\documentclass[useAMS,usenatbib,usegraphicx]{mn2e}
 
\title[Galaxy Structures at $z \sim 4-6$ in the Hubble UDF]{The Structures of Distant Galaxies - II: Diverse Galaxy Structures and Local Environments at $z = 4-6$; Implications for Early Galaxy Assembly}
\author[Conselice \& Arnold]{ Christopher J. Conselice$^{1}$\thanks{E-mail:
conselice@nottingham.ac.uk}, Jessica Arnold$^{2}$ \\
$^{1}$University of Nottingham, School of Physics \& Astronomy, Nottingham, NG7 2RD UK \\
$^{2}$California Institute of Technology, Pasadena, CA 91125}
\def\deg{$^{\circ}\,$}
\def\solm{M$_{\odot}\,$}

\def\deg{$^{\circ}\,$}
\def\solm{M$_{\odot}\,$}

\def\casgm20{CAS-G-M$_{20}\,$}
\def\m20{M$_{20}\,$}
\begin{document}

\date{Accepted ; Received ; in original form}
\pagerange{\pageref{firstpage}--\pageref{lastpage}} \pubyear{2002}

\maketitle

\label{firstpage}

\begin{abstract}
 
We present an analysis of the structures, sizes, star formation
rates, and local environmental properties of galaxies at $z \sim 4 - 6$ 
($\tau_{\rm universe} < 2$ Gyr), utilising deep Hubble Space Telescope 
imaging of the Hubble Ultra Deep Field.  
The galaxies we study are selected with the Lyman-break drop-out 
technique, using galaxies which are B-,V-, and $i-$drops, which effectively 
selects UV bright starbursting galaxies between $z = 4$ and $z = 6$.   
Our primary observational finding is that  starbursting galaxies at 
$z > 4$ have a diversity in structure, with roughly 30\% appearing distorted
and asymmetric, while the majority are smooth and apparently undisturbed
systems.  We utilize several methods to compute the inferred assembly 
rates for these distorted early galaxies including utilising the CAS system
and pair counts.  Overall, we find a similar fraction of galaxies which
are in pairs as the fraction which have a distorted structure.   Using the
CAS methodology, and our best-estimate for merger time-scales, we find that
the total number of inferred effective mergers for M$_{*} > 10^{9-10}$ \solm 
galaxies  at $z < 6$ is $N_{\rm m} = 4.2^{+4.1}_{-1.4}$.  The more common
symmetrical systems display 
a remarkable scaling relation between the concentration of
light and their half-light radii, revealing the earliest known galaxy 
scaling relationship, and demonstrating that some galaxies at $z > 4$
are likely in a relaxed state.   Systems which are asymmetric do not display a
correlation between size and half-light radii, and are generally larger
than the symmetric smooth systems.  The time-scale for the formation of
these smooth systems is 0.5-1 Gyr, suggesting that most of these galaxies are 
formed through coordinated very rapid gas collapses and
star formation over a size of 1-2 kpc, or from merger events at
$z > 10$.  We finally investigate the relation
between the UV measured star formation rates for these galaxies and their
structures, finding a slight correlation such that more asymmetric systems
have slightly higher star formation rates than symmetric galaxies.

\end{abstract}


\section{Introduction}

Distant galaxies are now routinely detected and studied out to 
$z \sim 6-7$ through
a variety of approaches and techniques. These young galaxies represent 
the empirical limit to our current understanding of the formation of the 
first galaxies in the universe.     The most common methods
for locating these systems include: deep imaging, usually through
Hubble Space Telescope surveys (e.g., Dickinson et al. 2004; Stanway
et al. 2003), narrow-band filter imaging searches (e.g., Rhoads et al. 2003;
Kashikawa et al. 2006), 
and deep blind spectroscopic surveys (e.g., Stark et al. 2007).  Of
interest to this paper, during the last
few years, dozens of high redshift, $z > 5$, galaxies have been identified 
in deep Hubble Space Telescope imaging, including the Hubble Ultra Deep Field 
(Beckwith et al. 2006; hereafter the UDF), and the Great Observatories
Origins Deep Survey (GOODS; Giavalisco et al. 2004a), among others.

Both the UDF, GOODS, and other surveys have been used to locate galaxies at 
$z > 4$, measuring and constraining, among other things, the global 
star formation rate history out to $z \sim 6$ (e.g., Giavalisco et al. 
2004b; Bunker et al. 2004;  Bouwens et al. 2006).    It appears from 
these observations that the star formation rate at early 
epochs is high, although not as high as the star formation rate peak at 
$z = 1.5 - 3$ (e.g., Bouwens et al. 2007). Based on {\em Spitzer} imaging many of 
these galaxies are found to be quite 
massive, with stellar masses M$_{*}$ $> 10^{9-10}$ \solm\, (e.g., Yan et al. 2005; Eyles 
et al. 2007), as well as having stellar population ages suggesting the onset 
of their formation was several hundred million years earlier (e.g., 
Yan et al. 2005; Eyles et al. 2007; Stark et al. 2007).   To probe 
higher redshifts, and to learn more through spectroscopy about the physical 
nature of 
already confirmed $z > 5$ galaxies will likely require the next 
generation of  20-30 meter telescopes, and the 
{\em James Webb Space Telescope}.   Currently we do not know when
the first galaxies formed, nor do we know much about the physical
processes driving the formation of these earliest galaxies.

A parameter space of these galaxies, which however has not been explored in 
any detail, is the structures and sizes of $z > 4$ galaxies (e.g., Ferguson
et al. 2004; Bouwens et al. 2004; Ravindranath et al. 2006; Hathi
et al. 2007).  Nor has it yet been determined whether structural features
of $z > 4$ galaxies can reveal how the earliest
galaxies are forming, and their connection to lower redshift systems 
(e.g., Conselice et al. 2008).  While galaxy images, even when
studied with HST, do not provide the same kind of information
as spectroscopy, imaging is in some cases more powerful than spectroscopy
for determining how a galaxy population is evolving.  The
main reason is that with deep Hubble imaging more galaxies can
be studied in a given population than can possibly have spectroscopy
reliably measured. In the case of individual $z > 5$ galaxies, very little
information beyond a redshift is provided by spectroscopy, although stacked
spectra can provide more detailed information about gas outflows and
other physical processes (e.g., Vanzella et al. 2009).  Imaging on the other
hand provides unique information for as many resolved galaxies
that can be imaged with a high enough signal to noise 
(e.g., Conselice 2003).

The extreme depth and high resolution of the UDF  data 
allows us to probe the internal structures of these first galaxies.
Using known structural properties of $z < 3$ galaxies (e.g., Conselice 2003;
Conselice et al. 2008; Lotz et al. 2008a) we can potentially determine the 
physical methods whereby these first galaxies formed, and their connection
to lower redshift systems.   We take a general approach 
to this problem in this paper by examining the qualitative and quantitative 
structures, and the incidence of likely merging pairs of Lyman-break 
galaxies at $z > 4$.  Our goal is to determine if the formation mechanisms for
these very early galaxies can be studied using their resolved ultraviolet
structures, and what we can learn about the earliest formation modes, and
its history, by using these features.

Furthermore, understanding the structural properties and 
evolution of these first galaxies has 
profound consequences for cosmology and
structure formation.  Theories of galaxy formation based on Cold Dark Matter
and a $\lambda-$dominated universe
predict that galaxies form hierarchically, and this is especially true
at these very early epochs.  In Cold Dark Matter dominated models, 
gas rich galaxies at $z > 4$ collide and merge to form more massive systems, 
while at the same time triggering star formation, and the growth and 
assembly of central black holes.  The relative role of discrete merger
events as opposed to gas accretion from the intergalactic medium is
debatable, with some current models predicting that accretion is
the dominant method for building stellar mass at high redshifts 
(e.g., Keres et al. 2005).  There is now strong observational evidence
however that the merger process occurs at redshifts $z < 3$ based on observations
of the Hubble Deep Field North and South, and the Hubble Ultra Deep Field
(Conselice et al. 2003a; Conselice et al. 2004; Lotz et al. 2008a;
Conselice et al. 2008).  These earlier results found that  the merger 
fraction and rate increase with increasing redshift,  especially for the 
most massive and luminous galaxies, out to $z \sim 3$ (Paper I; Paper III;
Bluck et al. 2009).  

Observationally, the role of mergers in forming galaxies at $z > 3$ is
largely unconstrained.  The Hubble Ultra Deep Field is
perhaps our best opportunity to address this question within the
next decade due to its unprecedented depth.  There are several
issues however that we must confront when utilising even very
deep Hubble Space Telescope data for
this analysis at extreme redshifts, $z > 4$, where structural and
morphological analyses have not previously been performed. These issues
include, but are not limited to: the extreme cosmological surface brightness 
dimming, resolved galaxies, and morphological k-corrections.  
We discuss all of issues in this paper, as well as in Conselice et al. (2008) 
for galaxies at $z < 3$ within the UDF.

Overall, we examine in several ways the structures of $z > 4$
galaxies as seen within the Hubble UDF field.  This includes: 
investigating their apparent and quantitative structural features, the 
sizes of these early galaxies, the incidence of systems in pairs, and the 
relation of these quantities with the ongoing star formation rate.  We 
interpret these observations to imply that there is a diverse, and likely 
rapidly changing, formation history for $z > 4$ galaxies.    We conclude
that a substantial fraction of $z > 4$ LBGs are possibly in a merger or
rapid assembly phase, but we also find that a significant number of systems 
appear relaxed, with a well-defined correlation between the 
concentration of light and their half-light radii.  We also give a 
description  for how best to analyse extremely faint galaxy data sets 
which will be useful for future analyses using faint resolved galaxy images.

This paper is organised as follows: \S 2 includes a discussion of the 
data sources we use in this paper, and the sample selection, \S 3 is a 
description of our morphological and structural analyses methods, \S 4 gives
a detailed investigation into our errors and systematics associated with
measuring structure on these galaxies, 
\S 5 presents our analysis, \S 6 is a discussion of our results and their
implications, and finally \S 7 is our summary and conclusions.  Readers
interested in skipping the technical details of our analysis are advised
to read \S 3.1 \& \S 3.2, and from \S 5 onwards.   
We use a standard cosmology of H$_{0} = 70$ km s$^{-1}$ Mpc$^{-1}$, and 
$\Omega_{\rm m} = 1 - \Omega_{\lambda}$ = 0.3 throughout.

\section{Data and Sample Selection}

\subsection{Data}

The primary data source used in this paper are the ACS and NICMOS
imaging of the Hubble Ultra Deep Field (Thompson et al. 2005;
Beckwith et al. 2006; UDF). Descriptions of the UDF survey are
included in e.g., Beckwith et al. (2006), Coe et al. (2006) and Conselice
et al. (2008), and references therein.
The UDF ACS uses the same filter set as the GOODS survey, which are the F435W 
(B$_{435}$), F606W (V$_{606}$), F775W ($i_{775}$), and F850L ($z_{815}$) 
bands. The UDF programme used 400 orbits of Hubble imaging for a
total exposure time of just under 1 Msec.
The field of view of the ACS image for the UDF is 11 arcmin$^{2}$, and is 
located within the GOODS-South field.   The central wavelengths of the 
filters we use, and their full-width at half-maximum,
are: F435W (4297, 1038 \AA), F606W (5907, 2342 \AA), 
F775W (7764 1528 \AA), F850L (9445 1229 \AA) The limiting 
magnitude for point sources is m$_{\rm AB} \sim 28.7$ in the $z-$band at
10 $\sigma$ within the UDF images using a 2\arcsec aperture, 
making the UDF easily the deepest optical imaging taken to date.  The
other bands have similar 10 $\sigma$ depths. Further 
details concerning the UDF imaging are 
presented in Beckwith et al. (2006) and Coe et al. (2006).  All
structural analyses in this paper are done using the $z-$band
ACS imaging.

Since we examine very faint and small galaxies within the
UDF, we discuss briefly some of the features of the data reduction, and
the image quality of the ACS imaging.  First we note that we only use
the $z-$band F850L image for our analysis, to obtain the reddest 
wavelength possible without using NICMOS imaging which has a PSF size
between 0.22-0.3\arcsec.   The data products we use in this analysis
originate mostly from the reduction
and cataloguing from Beckwith et al. (2006) and Coe et al. (2006).  
Beckwith et al. (2006) provides a detailed description of the data reduction 
and data acquisition procedures for the UDF, including the justification for 
the depth, field selection, and filter choices.  Beckwith et al. (2006)
also describes in some detail the data reduction and image quality tests which 
we summarise here.

The UDF was a major imaging campaign with the Hubble Space Telescope utilising
director's discretionary time.  The UDF data was acquired over a broken
time-span from September 24 2003 until January
15 2004, utilising various roll angles and dithering positions.  The roll
angles used to take the data: 40\deg, 44\deg, 310\deg and 314\deg, were 
selected to obtain a nearly square final image, unlike for
the GOODS imaging where various roll angles were used to optimise for
supernova searches (Giavalisco et al. 2004a).  Each ACS observation
of the UDF consisted of 2 orbits, divided into two exposures per orbit,
with typical single exposure times of 850-1200s (Beckwith et al. 2006).
In addition to different roll angles, the UDF images were dithered with
sub-pixel sized shifts between exposures, in an attempt to obtain higher
quality images through the drizzle process.

The dither pattern used in the UDF data acquisition was a four point
dither of integer pixels combined with half-integer offsets to both
increase the image quality, as well as to make easier the removal of bad
pixels and columns, cosmic rays, and other defects.  The resulting images 
were then combined, after
sky subtraction through a method outlined in Beckwith et al. (2006),
using the MultiDrizzle programme.  The main task of MultiDrizzle is to
produce for each exposure a rectified output image on a common
grid that are then all combined to create a single final image.  
The output pixel scale was set to 0.6 times the original ACS WFC scale of 
0.05\arcsec pixel$^{-1}$, for a final pixel scale of 0.03\arcsec pixel$^{-1}$ 
which allows Nyquist limited sampling of the point spread function.  The 
output image was created such that each pixel in the image maps from 
a single pixel in the input images. This reduces the amount of processing on 
the images themselves, while providing the highest resolution possible with a
limited amount of correlated noise.  Tests of the image quality of the
PSF based on stars reveals FWHM values of 0.089\arcsec in the $z-$band
(see \S 3.3 for a more detailed discussion of this.)

The initial photometry for the UDF galaxies were taken from the publicly 
available catalogs of the UDF from Beckwith et al. (2006).  This catalog and
photometry was obtained through the use of SExtractor, including the
use of the output segmentation maps, used later in the morphological
analysis.  The output from the SExtractor process produces a catalog of
photometry at various apertures, including total magnitudes, as well
as other photometric, and some size measurements.  Output from the
SExtractor process includes a so-called segmentation map which is a matched 
map, aligned with the original image, and is constructed such that each pixel 
in the segmentation map is mapped to either a single galaxy or to the 
background.  This is done to both define the galaxy for the photometry, as 
well as for the structural analysis we perform in this paper.  
The magnitudes we use are all based on this catalog, where we utilise the 
total mags from the SExtractor output which gives a magnitude in each of
the ACS filters we examine: BV$iz$, which are then used for the Lyman-break
galaxy selection.

\subsection{Sample}

Our sample selection matches that used in previous Lyman-break galaxy
surveys of the Great Observatories Origins Deep Survey (e.g., Giavalisco et 
al. 2004b; Dickinson et al. 2004; Ravindranath et al. 2006; Lee et al. 2006) 
which is designed to locate galaxies at $z \sim 4, 5$ and 6.  We utilise 
the so-called `drop-out' 
technique to find these galaxies, applying the criteria discussed in
Giavalisco et al. (2004b) and Dickinson et al. (2004).   These criteria are:

$$(B_{450} - V_{606}) \geq 1.2 + 1.4 \times (V_{606} - z_{850}),\, {\rm and} $$ 
$$(B_{450} - V_{606}) \geq 1.2,\, {\rm and}\, (V_{606} - z_{850}) \leq 1.2,$$

\noindent for B-drops, which are at $z \sim 4$.  For the V-drops (objects
at $z \sim 5$) we use the criteria:

$$[(V_{606} - i_{775}) > 1.5 + 0.9 \times (i_{775} - z_{850})], {\rm or}$$
$$[(V_{606} - i_{775}) > 2],\, {\rm and}\, (V_{606} - i_{775}) \geq 1.2,\, {\rm and}$$
$$(i_{775} - z_{850}) \leq 1.3.$$

\noindent Finally, for the $z \sim 6$ $i-$drops we use the criteria,

$$(i_{775} - z_{850}) \geq 1.3.$$

\noindent Although high$-z$ candidates have been selected and published in 
the UDF area by Yan \& Windhorst (2004), Bunker et al. (2004) and Beckwith
et al. (2006) we recompute
our own samples.  Using the selection criteria above we obtain candidate 
objects at redshifts $z \sim 4, 5, 6$ respectively, finding similar
surface densities of drop-outs as these previous works. We include in this 
list the 
spectroscopically confirmed systems published in Stanway et al. (2003), 
Dickinson et al. (2004), and Stanway et al. (2007), including
the redshifts compiled in Coe et al. (2006).  We also place
a magnitude limit of $z < 28.5$ on our initial selection for drop-outs.   
We remove any point
sources based on the stellaricity index calculated during the
SExtracting process. Images of the $z < 28$ $i-$drops are shown in Figure~1.

\begin{figure*}
%\vspace{5.5cm}
 \vbox to 135mm{
\includegraphics[angle=0, width=174mm]{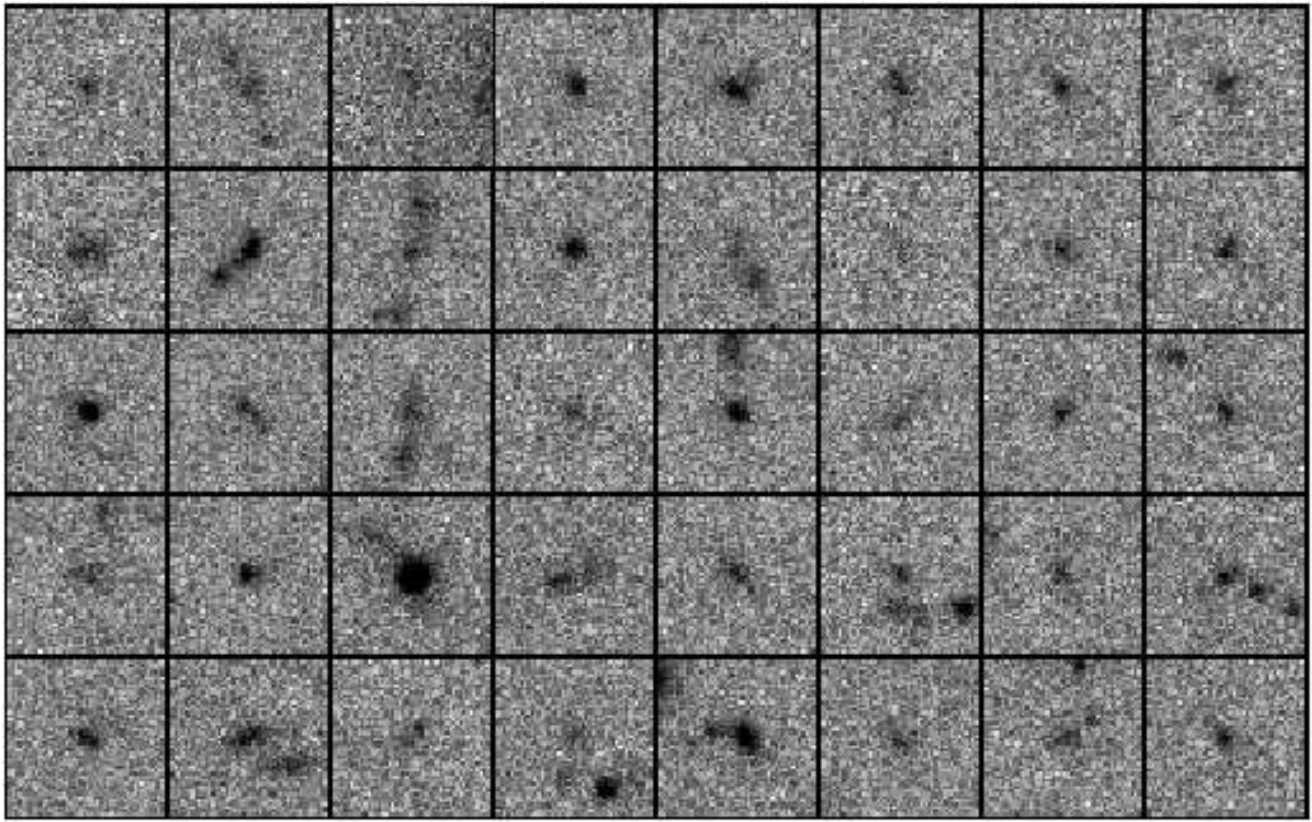}
 \caption{Mosaic of UDF $i-$drops at $z < 28$ that
remain after our purging (\S 2.2) within our sample as seen
within the $z-$band UDF imaging.  These are roughly
ordered from smoothest systems to more distorted ones from right to left,
top to bottom. The field of view
of each image is 2.3\arcsec, or 13 kpc a side, at $z \sim 6$. }
\vspace{10cm}
} \label{sample-figure}
\end{figure*}

We examined all of our candidate drop out systems by eye for classification
purposes, as well as to remove contamination, such as lower redshift 
galaxies and stars.  If a candidate drop-out
is detected in a band bluer than the drop-out band,
it is removed from consideration.   In the case of the
B-band drop-outs we use photometric redshifts and whether galaxies
are detected in the B-band itself to remove lower redshift
contamination, which we find to be very minor (e.g., Bunker et al. 2004).  
We carry out this purging and classification by examining all 
of the ACS wave-bands.
After carrying out these procedures we find within the UDF a total of 126 
i-drops, 137 V-drops and 320 B-drops.   For, in particular
the $i-$drops, there is the issue of contamination by lower-redshift
galaxies, particular from evolved ellipticals at $z = 1-2$, and from galactic
stars. As mentioned earlier, we remove any systems which are unresolved within
our drop-out list, which in the UDF is
only a few systems, which are all brighter than mag $z \sim 25$.  
In fact, based
on NIR colours, the contamination rate for $z > 25.6$ $i-$drops is roughly
1-2\%, much lower than the 1/3 contamination rate seen in the brighter
drop-out systems (e.g., Bunker et al. 2004; Dickinson et al. 2004).

Finally, we emphasis that we only use the $z-$band ACS imaging for our
morphological/structural/size measurements.  While the NICMOS data
sample longer rest-frame wavelengths for these galaxies, we do not
use this imaging as the NICMOS PSF is much larger than it is for ACS, making
it very difficult to impossible to use for morphological measurements.
We also apply a S/N cut of $> 10$ for our morphological analysis as
well as an overall magnitude limit of $z_{850} = 27.5$ (Figure~2) to
only include systems that are bright enough for structural analyses
(\S 4).  This leaves 69 B-drops, 43 V-drops and 21 $i-$drops for this
part of our analysis.

\section{Structure, Size, Morphology and Classification Methods}

In this section we describe our methodology for measuring the
various properties of our drop-out sample.  The procedure for
doing this is explained below, but can be summarised briefly.
Each drop-out is cut-out from the UDF image and examined in the
$z-$band by eye (in \S 2 we explain how we also examined these
galaxies in the other bands) and classified according to
the criteria in \S 3.1.  We then analysed these images through
the CAS code and method (e.g., Conselice 2003), which provides
measures of total radii (Petrosian radius), half-light radii,
fluxes within the total radii, as well as the CAS parameters
themselves.

\begin{figure}
%\vspace{5.5cm}
 \vbox to 110mm{
\includegraphics[angle=0, width=90mm]{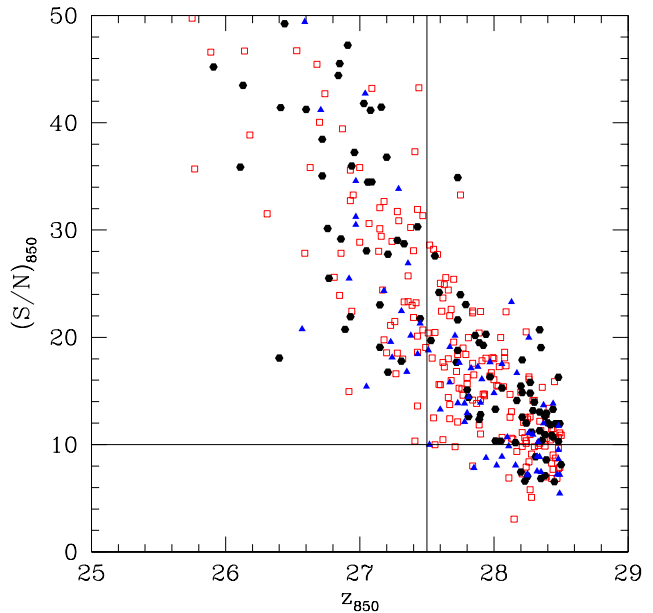}
 \caption{The distribution of S/N for our sample, as a
function of the observed $z_{850}$ magnitude.  The open
red boxes are for B-drops, the solid black symbols
are the V-drops, and the blue triangles are for
the $i-$drops in our sample.  The horizontal line
denotes our S/N limit and cut which we use throughout
this paper.   The vertical line is the magnitude limit
in which we select our sample for the CAS analysis. }
\vspace{3cm}
} \label{sample-figure}
\end{figure}

We use in this paper two methods for classifying our drop-outs. 
The first is a simple
examination in the $z_{850}$-band of the apparent structure and morphology
of each system.  This is not meant to be definitive, nor is any
physical meaning necessarily implied by these classifications.  The
second type of classification method involves a quantitative approach
using a revised CAS and Gini/\m20 methodology introduced in Conselice
et al. (2008), and described in more
detail in \S 3.2.  We describe both methods and their limitations below.

\subsection{Visual Typing}

The first part of our analysis involves examining every
drop-out in each of the four main UDF ACS bands, for various
purposes (\S 2). The process
for carrying this out involves cutting out into a postage
stamp-sized image each drop-out in the BV$iz$ bands and then
examining these images by eye to determine whether the galaxy
appears in that given band, and if so, what its visual
morphology is.  Figure~1 shows a $z < 28$ selected $i-$drop
sample. The large number of drops in the other bands, prohibits
showing the other samples.   We do all of our classifications
of types in the $z-$band for all drop-outs.  If a galaxy is detected 
and meets our analysis criteria, it is then classified into one of the 
following classes:

\begin{enumerate}

\item \noindent  Normal - The galaxy is resolved and appears
to have a normal, roughly symmetric, shape.

\item \noindent  Elongated - The galaxy appears elongated

\item \noindent  Neighbour - The galaxy is near another galaxy
which is also a drop-out in the same band; examples of these systems
are shown in Figure~3

\item \noindent  Unusual/asymmetric/peculiar - the galaxy appears unusual
in some way, typically asymmetric, with examples of these systems
shown in Figure~4

\begin{figure*}
%\vspace{5.5cm}
 \vbox to 150mm{
\includegraphics[angle=0, width=174mm]{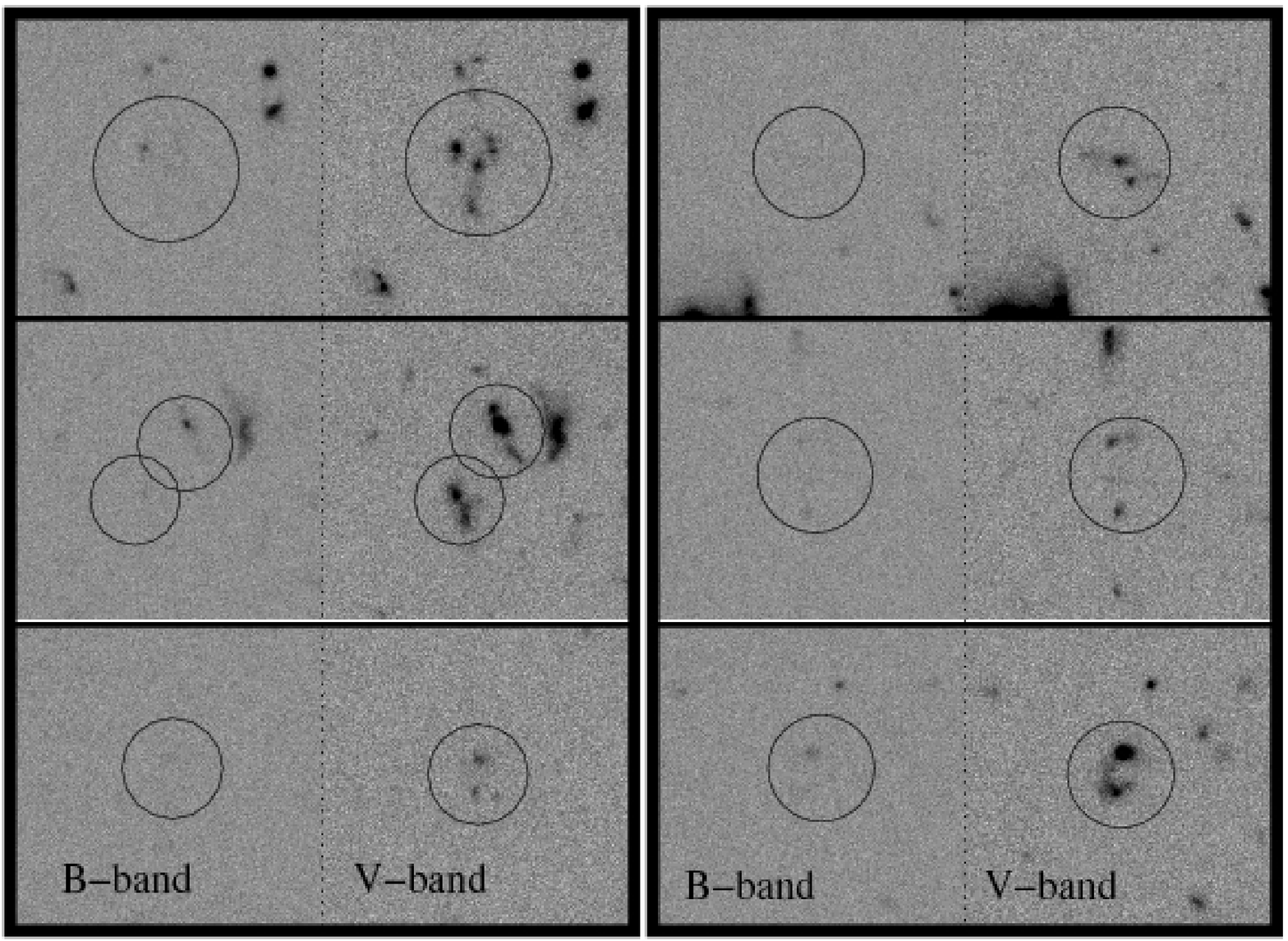}
 \caption{Examples of B-band Lyman-break drop-outs which are
in pairs, as found within the UDF.  Shown here are six
examples of these systems, with the right panel showing
the V-band image, while the left panel shows the B-band
image.  These systems very often show signs of tidal
distortions, such as extended low surface brightness
light, similar to nearby mergers and those systems
shown in Figure~4.  The circle on each image shows the
region in which the drop-out is present in the V-band,
and the corresponding location in the B-band.  The field
of view of each image is roughly 5\arcsec on a side, or
$\sim 35$ kpc at the redshift of these drop-outs.}
%\vspace{3cm}
} \label{sample-figure}
\end{figure*}

\begin{figure*}
%\vspace{5.5cm}
 \vbox to 150mm{
\includegraphics[angle=0, width=174mm]{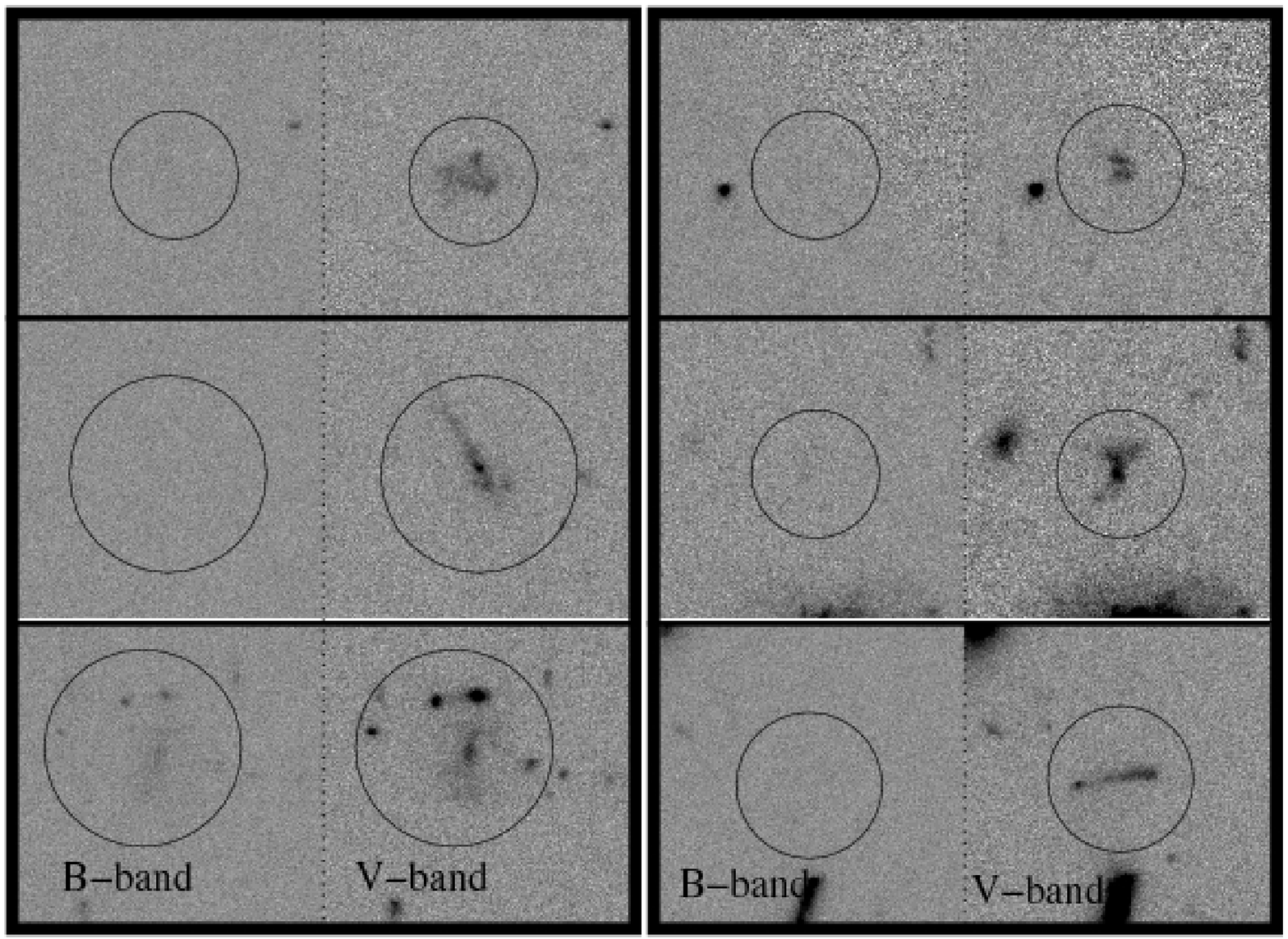}
 \caption{Examples of B-band Lyman-break drop-outs,
which are classified as peculiar, as found within the UDF.  
Shown in this figure are six examples of these systems, with 
the right panel showing the V-band image, while the left panel 
shows the B-band image.  These systems all show signs of tidal
distortions, such as extended low surface brightness
light, and perhaps tidal bridges between various galaxies.
The circle on each image shows the
region in which the drop-out is present in the V-band,
and the corresponding location in the B-band. The field
of view of each image is roughly 5\arcsec on a side, or
$\sim 35$ kpc at the redshift of these drop-outs.}
} \label{sample-figure}
\end{figure*}

\item \noindent  Star-like - The object appears unresolved and
very compact, such as a star.

\end{enumerate}

\subsection{The Extended CAS Structural Analysis}

We use the CAS (concentration, asymmetry, clumpiness) parameters to measure
the structures of our $z > 4$ galaxies quantitatively.  We include in our
analysis the measurement of the Gini and \m20 parameters (e.g.,
Lotz et al. 2008a).  The CAS/Gini/\m20 parameters  are a 
non-parametric method for measuring the forms of galaxies on resolved 
CCD images (e.g., Conselice 1997; Conselice et al. 2000a; Bershady et al. 2000; 
Conselice et al. 2002;  Conselice 2003; Lotz et al. 2008a).   The
basic idea behind these parameters is that galaxies have light distributions
that reveal their past and present formation modes (Conselice 2003). This
system is well calibrated at $z \sim 0$, and to a lesser degree at
$z < 1$, but its use and applicability at $z > 4$ remains untested
until this paper.

One of the major benefits of the CAS system is that well-known galaxy 
types in the nearby universe fall into
specific regions of CAS parameter space.  For example, the 
selection $A > 0.35$ locates systems which are nearly all major
galaxy mergers in the nearby universe (e.g., Conselice et al. 2000b; 
Conselice 2003; Hernandez-Toledo et al. 2005; Conselice 2006a).
In addition to the classic CAS parameters, we also investigate
the use of the similar Gini and M$_{20}$ parameters (Lotz et al. 2008a) for
understanding the morphologies of the UDF galaxies.  Our method is
the same as used in Paper I (Conselice et al. 2008).
A brief description of the parameters we use in this analysis
 is provided below.

The way we measure these structural parameters on the UDF images 
varies slightly
from what has been done earlier in the Hubble Deep Field, and early GOODS 
studies (e.g., Conselice et al. 2003a; Mobasher et al. 2004;
Conselice et al. 2004). Our basic
procedure is to cut out each galaxy in our sample into a smaller image 
from which
the entire analysis is done.  The same part of the weight map and
segmentation map is cut out as well.  Next the code measures the radius in 
which the parameters are computed.   We use the sizes measured through
the CAS code, namely Petrosian radii, and the half-light radii, throughout
this paper.  The total flux is then also measured within this Petrosian
total radius. 

\subsubsection{Measured sizes}

The radius we use for all of our indices is the Petrosian radii, 
which is the
radius defined as the location where the surface brightness at a 
radius is 20\% of the surface brightness within that radius (e.g.,
Bershady et al. 2000; Conselice 2003).  As described in Bershady
et al. (2000), for most galaxy profiles, this Petrosian radius will
contain 99\%, or nearly all of the light in a galaxy.  This radius
has also been used in nearly all structural analysis studies at
high redshift, and is even becoming a standard radius for
measuring galaxy sizes in the nearby universe, such as within the
Sloan Digital Sky Survey (e.g., Graham et al. 2005).
 
We use circular apertures for
our Petrosian radii, and for our quantitative parameter estimation.  The
reason we do not use elliptical or more complicated radii is to avoid 
ambiguity produced through assumptions about the shape of
our galaxy sample.  Furthermore, many of our galaxies have such irregular
and peculiar structures, that anything but a single circular aperture would
be too complex to interpret through our structural methods.  We begin
our estimate of the galaxy centre for the radius measurement at
the centroid of the galaxy's light distribution.  Through modelling
and various tests, we have previously shown that the resulting radii do not
depending critically on the exact centre, although the CAS and other
parameters do (Conselice et al. 2000a; Lotz et al. 2008a).  The 
Petrosian radius we use to measure our parameters is defined by:

$$R_{\rm Petr} = 1.5 \times r(\eta = 0.2),$$

\noindent where $r(\eta = 0.2)$ is the radius where the surface
brightness is 20\% of the surface brightness within that radius, or,

$$\eta = \frac{I(R)}{<I(r)>} = 0.2.$$

\noindent  This follows the suggested form given by Kron (1995), where
the value of $\eta(r)$ is equal to unity at the centre of a galaxy, and
goes to zero at large galactic radii.
Typical Petrosian radii for our sample typically range from 
0.2-0.6\arcsec, while the half-light radii vary between 0.1-0.3\arcsec
(see \S 3.3).

The CAS code furthermore measures the flux within this Petrosian radius,
where we measure our morphological and structural parameters. To test
how well we are measuring the total light from these galaxies we compare
the CAS measured magnitude to the SExtractor measured total magnitudes
discussed in \S 2.1 in Figure~5.  Figure~5 shows that there is a good
relation between the two measured magnitudes, with the average differences,
and 1 $\sigma$ variation on these differences for the various drop-outs: 
$\delta$mag = 0.03$\pm$0.13 for the B-drops, $\delta$mag = 0.03$\pm$0.12 
for the V-drops, and $\delta$mag = 0.005$\pm$0.13 for the $i$-drops. These
differences are such that the SExtractor magnitudes are slightly brighter
than the CAS measured ones by $\sim$ 0.5-3\%. Most
of the scatter is produced by the faintest galaxies, most of which we
do not use in our analysis.

\begin{figure}
%\vspace{5.5cm}
 \vbox to 140mm{
\includegraphics[angle=0, width=90mm]{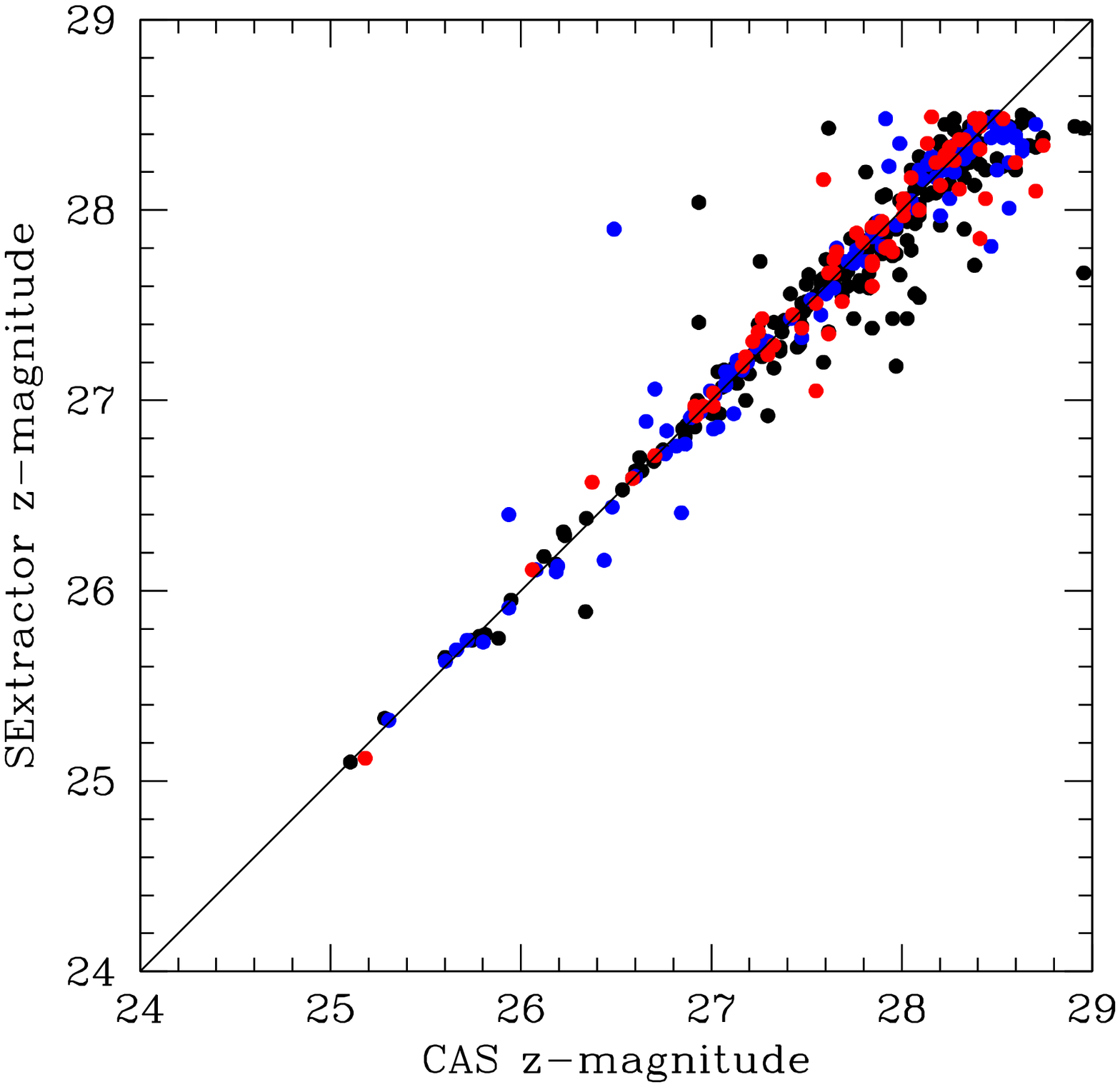}
 \caption{The relation between the measured $z$-band magnitudes as measured
with SExtractor, as published and catalogued in Beckwith et al. (2004),
and those measured through the CAS method and code within the
$\eta = 0.2$ radius as discussed in \S 3.2.  The black points are
for the B-drops, the blue points are for the V-drops, and the
red points are the $i-$drops.  We find a very good relation between
these two methods of measuring the magnitudes for our sample (\S 3.2.1).  }
%\vspace{3cm}
} \label{sample-figure}
\end{figure}

\begin{figure*}
%\vspace{5.5cm}
 \vbox to 140mm{
\includegraphics[angle=0, width=174mm]{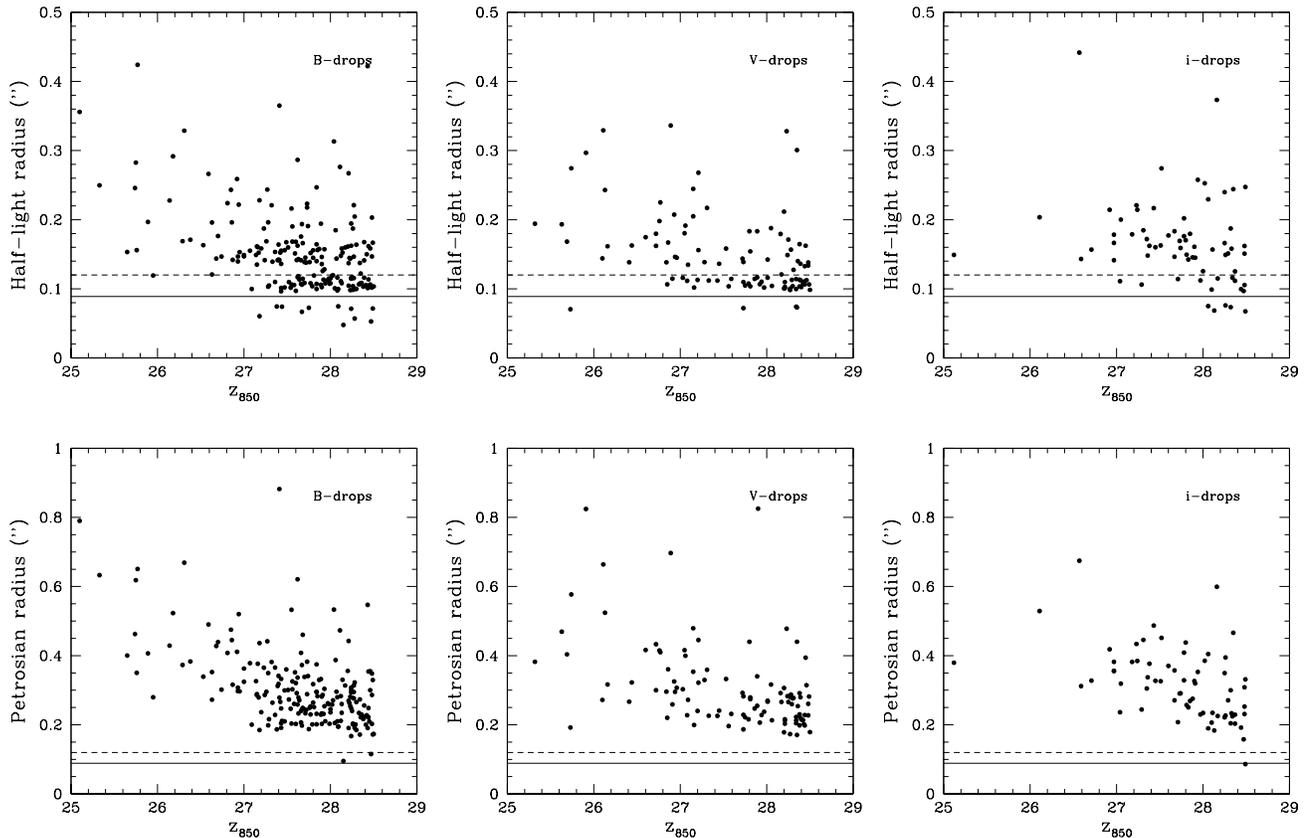}
 \caption{Top: The relationship between the half-light radius
of our galaxies, as measured in arcsec, vs. the magnitude
of our sample in the $z-$band.  The solid horizontal line
is the measured ACS PSF FWHM within the $z-$band, and the dashed
horizontal line is the undrizzled ACS PSF FWHM. Bottom: 
similar to the top panels, but displaying
the Petrosian radius as a function of magnitude for the
same galaxies.}
%\vspace{3cm}
} \label{sample-figure}
\end{figure*}

Another important issue, especially for faint galaxies seen in the
UDF, is how to account for background light and noise. For faint galaxies there
is a considerable amount of noise from the sky, which must
be accounted for.  Through various test we conclude that the proper
way to correct parameters for the background requires that the selected
background area be near the object of interest. This is only an issue
for faint galaxies, and for galaxies imaged on large mosaics which
have a non-uniform weight map, and whose noise characteristics vary
across the field.  By using a background near each object we alleviate these
issues as the noise properties do not vary significantly over 
$\sim 0.5 - 1$ arcmin, where the galaxy and the
background area are selected.  We review below how the CAS and Gini/\m20
parameters are measured. For more detail see Bershady et al. (2000),
Conselice et al. (2000a), Conselice (2003) and Lotz et al. (2008a).

\subsubsection{Asymmetry}

The asymmetry of a galaxy is measured by taking an original galaxy 
image and rotating it by 180 degrees about its centre (defined below), and then
subtracting the two images (Conselice 1997). Within this method 
there are corrections done for background, and radius, which are
explained in detail in Conselice et al. (2000a). Briefly, the most important
correction for the asymmetry index is the background light, and the
noise within this.  Furthermore, we measure the asymmetry out to the
Petrosian radius, although other similar radii give very similar results (e.g.,
Conselice et al. 2000a). The centre for rotation is decided by an iterative
process which finds the location of the minimum asymmetry.  The formula
for calculating the asymmetry is given by:

\begin{equation}
A = {\rm min} \left(\frac{\Sigma|I_{0}-I_{180}|}{\Sigma|I_{0}|}\right) - {\rm min} \left(\frac{\Sigma|B_{0}-B_{180}|}{\Sigma|I_{0}|}\right)
\end{equation}

\noindent Where $I_{0}$ is the original image pixels, $I_{180}$ is the image
after rotating by 180\deg from each estimated centre.  The background 
subtraction using light from a
blank sky area, called $B_{0}$, are critical for this process, and must 
be minimised in the same way as the original galaxy itself.  A lower value 
of $A$ means that a galaxy has a 
higher degree of rotational symmetry which tends to be found in
elliptical galaxies in the nearby universe.
Higher values of $A$ indicate an asymmetric light distribution, which are 
usually found in spiral galaxies,  or in the more extreme case, merger 
candidates (e.g., Conselice et al. 2000a; Conselice 2003).

\subsubsection{Concentration}

Concentration is a measure of the intensity of light contained within a 
central region, in comparison to a larger region in the outer-parts of a
galaxy.  There are various ways to measure the light concentration in a 
galaxy, with the most robust and reliable method being taking the ratio of
two radii which contain a certain fraction of the galaxy's light (e.g.,
Graham et al. 2001).  These two light fraction radii should differ
enough to ensure that galaxies of different types can be
distinguished, but there are limits to how small or large a flux
radius should be used.  The reason is that the radius cannot be too
small or else it will be confused with the PSF, and cannot be too large
as it will be hard to measure the `total' light due to sky subtraction
errors.  Bershady et al. (2000) investigated several forms for the
concentration index, and found that the C$_{28}$ index is the most
robust for small galaxies, as well as providing the most dynamic
range for separating galaxies of different types.  We use this
index in this paper.

The exact definition we use to measure light concentration is
the ratio of two circular radii which 
contain 20\% and 80\% ($r_{20}$, $r_{80}$) of the total galaxy flux,

\begin{equation}
C = 5 \times {\rm log} \left(\frac{r_{80}}{r_{20}}\right).
\end{equation}

\noindent   A higher value of $C$ indicates that a larger amount of light 
is contained within a central region.   This
particular measurement of the concentration correlates well with 
the mass and halo properties of galaxies in the nearby
universe (e.g., Bershady et al. 2000; 
Conselice 2003).  

One major issue we must confront in this paper is the fact that
the inner radius of r$_{20}$ is likely often smaller than the PSF
itself, and therefore we are potentially measuring incorrect
concentration index values.  We address this issue in
detail in \S 4.

\begin{figure}
%\vspace{5.5cm}
 \vbox to 120mm{
\includegraphics[angle=0, width=85mm]{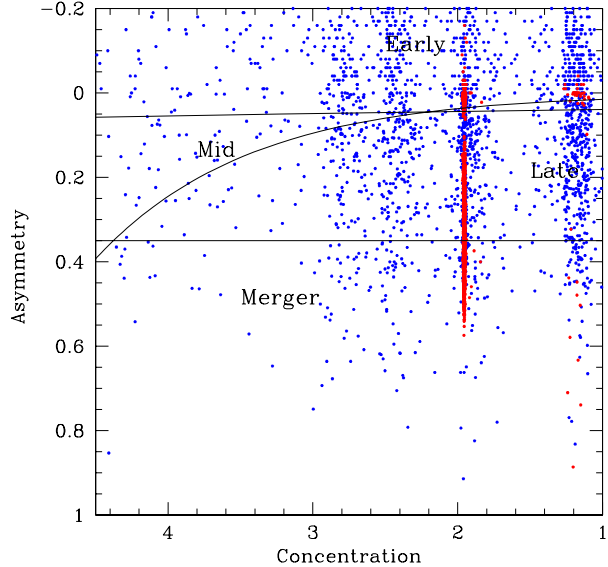}
 \caption{The concentration-asymmetry diagram for
faint galaxies in the COSMOS field (blue points) and
objects identified as stars within the same field (red points).  As
can be seen, the stars are found within a well defined region in
this plot and all contain a 
low-concentration, with a range of asymmetry values, while
the galaxies display a wide-range of both values.  }
%\vspace{3cm}
} \label{sample-figure}
\end{figure} 

\subsubsection{Clumpiness}

The clumpiness ($S$) parameter is used to describe 
the fraction of light in a galaxy which is contained in clumpy light
concentrations.   Clumpy galaxies have a relatively large amount of
light at high spatial frequencies, 
whereas smooth systems, such as elliptical galaxies contain light at low 
spatial frequencies. Galaxies which are undergoing star formation tend to 
have very clumpy structures, and high $S$ values.  Clumpiness can be 
measured in a number of ways, the most common method used,
as described in Conselice (2003) is,

\begin{equation}
S = 10 \times \left[\left(\frac{\Sigma (I_{x,y}-I^{\sigma}_{x,y})}{\Sigma I_{x,y} }\right) - \left(\frac{\Sigma (B_{x,y}-B^{\sigma}_{x,y})}{\Sigma I_{x,y}}\right) \right],
\end{equation}

\noindent where, the original image $I_{x,y}$ is blurred to produce 
a secondary image,  $I^{\sigma}_{x,y}$.  This blurred image is
then subtracted from the original image leaving a 
residual map, containing only high frequency structures in
the galaxy (Conselice 2003). To quantify this, we normalise the
summation of these residuals by the original galaxy's total light, and
subtract from this the residual amount of sky after smoothing
and subtracting it in the same way.  The size of the smoothing kernel 
$\sigma$ is
determined by the radius of the galaxy, and we use $\sigma = 0.2 \cdot 1.5
\times r(\eta = 0.2)$ (Conselice 2003), although other smoothing
scales are possible.  Note that the centres of galaxies are
removed when this procedure is carried out (e.g., Conselice 2003).  
We ultimately do not use the clumpiness index in this paper in
any extensive way due to the low dynamic range of values provided
by our systems due to their smaller sizes and faintness.

\subsubsection{Gini and \m20 Coefficients}

The Gini coefficient is a statistical tool originally used to 
determine the distribution of wealth within a population, with higher values 
indicating a very unequal distribution (Gini of 1 would mean all wealth/light 
is in one person/pixel), while a lower value indicates it is distributed 
more evenly amongst the population (Gini of 0 would mean everyone/every pixel 
has an equal share).   The value of G is defined by the Lorentz curve 
of the galaxy's light distribution, which does not take into 
consideration the spatial positions of pixels.  Each pixel is ordered by its
brightness and counted as part of the cumulative distribution (see
Lotz et al. 2008a).   

The \m20 parameter is a similar parameter to the concentration in that it 
gives a value that indicates 
whether light is concentrated within an image; it is however calculated 
slightly differently from C and Gini.  The total moment of light is calculated by summing the 
flux of each pixel multiplied by the square of its distance from the centre.  
The centre is the location where \m20 is minimised
(Lotz et al 2008).  The value of \m20 is the moment of the 
fluxes of the brightest 20\% of light in a galaxy, which is
then normalised by the total light moment for all pixels (Lotz
et al. 2008a).

The main differences between \m20 and $C$ are due to the moments in
\m20 which depend on the distance from the galaxy centre. The value of
\m20 will therefore be more affected by spatial variations, and also the 
centre of the galaxy is again a free parameter.  This makes it more 
sensitive to possible mergers.

\subsection{Effects from the ACS PSF}

Because many of our sample galaxies are small we must examine the ACS point 
spread function in the UDF and compare
its profile with the sizes of our galaxies in detail.  The ACS 
PSF has been well described in several papers, such as Sirianni et al. (2005)
and Rhodes et al. (2007), including studies that
carefully analyse ACS for use within galaxy lensing studies 
(Rhodes et al. 2007).  As is well known, the ACS PSF is affected by 
optical abberations and geometric distortions, as well as by blurring 
from charge diffusion from neighboring pixels due to sub-pixel variations.  
The PSF quality is also affected by the jitter during the observations 
themselves.  Due to the large number of images, and careful dither
patterns (\S 2.1), these issues are perhaps better minimised in this
field than in any other ACS survey.

Simulations with {\tt tinytim}, and observations of stars themselves within
deep imaging surveys show that the FWHM of the PSF for stars imaged
within the wide-field camera of ACS is roughly 0.12\arcsec\, after 
convolving the intrinsic PSF width of 0.085\arcsec\, with the
detector pixels. Image tests within the UDF using stars, find that
the FWHM of these stars are on average: 0.084\arcsec\, in the B-band,
0.079\arcsec\, in the V-band, 0.081\arcsec\, in the $i-$band and
0.089\arcsec\, in the $z-$band (Beckwith et al. 2006).  The scatter on these 
measurements is
roughly 1-2 mas, and agree with the expected values after considering
the initial convolution from discrete sampling in the 0.05\arcsec\, pixels,
and PSF smearing from the charge diffusion kernel in adjacent pixels,
and convolution by the 0.03\arcsec\, output pixel size (Beckwith
et al. 2006).   This PSF FWHM furthermore does not change significantly
between stars in the centre of the UDF and those towards the edges.

To determine the effects of the PSF on our analysis we first 
investigate the size distribution
of our galaxies in arcsec, as a function of magnitude.  This comparison
with the size of the ACS PSF is shown in Figure~6.  As can be seen, 
we find that, particularly at
faint magnitudes, there are galaxies with half-light radii which are slightly
smaller than the non-dithered WFC PSF.  
These galaxies are all fainter than $z = 27$.  However,
nearly all of our galaxies have diameters larger than the non-dithered ACS PSF.
Furthermore, only a handful of our galaxies have half-light radii
which are smaller than the measured PSF in these images (solid line
on Figure~6), and very few brighter than our ultimate analysis
limit of $z = 27.5$.  We also show in the bottom panels of Figure~6 the relation
between the Petrosian radius (\S 3.2) and the $z-$band magnitude.  This
demonstrates that all of our galaxies have total radii which are often
much larger than the size of the ACS PSF, both before or after drizzling.

There are a few other ways in which we determine the PSF effects
on our images and analysis. The first, as discussed above, is to 
determine the 
measured sizes of our galaxies compared to the size of the PSF.    The
other method is to measure the same parameters using the same code and
conditions on stars in our fields, and to determine how these stars,
which are effectively giving us a measure of the diversity in how the point 
spread function is sampled, behave in the various diagnostic methods we 
utilise.   Since the UDF has a limited field of view, and was designed to 
avoid stars, we utilise stars
found within the COSMOS field (see Paper III; Conselice et al. 2009) 
for later comparisons when discussing this aspect.

\section{Limits, Error Distributions and Systematics}

As this is the first major study to investigate in detail the morphological
and structural properties of $z > 4$ galaxies, it is important to understand
the sources of error and biases that are present when we examine distant 
galaxies.    These include effects from redshifts, both due to decreased
signal to noise and resolution, but also importantly, we must understand
and account for effects from the point spread function from ACS, as many
galaxies have half-light radii similar to, or just slightly larger than
the FWHM of the PSF (\S 3.3)

As mentioned earlier, there are a few other major problems that must be 
dealt with
when examining galaxies at these redshifts.  These include: cosmological 
surface brightness dimming, and morphological $k-$corrections,
as well as  resolution vs. galaxy size. Since higher redshift
galaxies are smaller on average than those at lower redshifts (e.g., Ferguson
et al. 2004; Trujillo et al. 2007; Buitrago et al. 2008) it is possible 
that even within the ACS
UDF image we are not imaging deep enough to see the
full structures of our $z > 4$ galaxies.  The long exposure times of the
UDF, and the use
of ACS, alleviates some of these issues, which we explore in 
depth.  

There are several ways in which we determine the systematic
and random errors on our quantitative parameter measurements.  One
way is by simulating lower redshift galaxies to higher
redshifts to determine how different parameters change purely
due to redshift effects.  This method was pioneered in Conselice
et al. (2000a), Bershady et al. (2000) and Conselice (2003), and remains 
an effective way to
account for the effects of redshift.  Another method to determine
our likely error distribution is to use the data
themselves to determine the likely systematics and how they are
distributed, and whether they are accounted for by our measured
errors.

\subsection{Simulations}

An important method for determining any systematic biases
when comparing galaxies at various redshifts  is to quantify how 
measures of galaxy size and structure change when a nearby
galaxy is placed at
larger distances.   The measurement of galaxy sizes and structure 
will change as the same galaxy becomes more distant, for example
in a simulation, due to decreased resolution and more noise due
to a lower measured flux.  For distant galaxies at $z > 1$, where
the angular size distance does not change much in our
cosmology, resolution for a galaxy of a given size and brightness is 
not important beyond the fact that it produces a lower surface brightness.  
However, if galaxy sizes decrease with time, as LBGs are thought to
do (e.g., Ferguson et al. 2004), then resolution can have an important
effect on the measured sizes and structural parameters.

The effects of distance on structural parameters, such as CAS, 
have been discussed in detail, and are well calibrated
in previous papers for galaxies at $z < 3$.  The typical 
way to carry out these types of analyses is to take galaxies at
$z = 0$ (Conselice et al. 2000a; Bershady et al. 2000; Conselice 2003; 
Conselice et al. 2003b) or 
$z = 1$ (Conselice et al. 2005) and simulate the same galaxies to how
they would appear at higher redshifts within our observational
conditions and parameters (i.e., UDF exposure time, HST aperture, 
ACS PSF, etc.), sans any morphological
k-correction.  One problem with this approach
is that it assumes that the galaxy which is being simulated is intrinsically
similar, or that the size and structure behave in a similar way, to the
higher redshift galaxy population which is being studied. However,
it is very unlikely that the galaxies we see at $z > 4$
are similar to galaxies at $z < 1$ due to the 
structure-redshift relation (Conselice et al. 2005), as well as due
to differing stellar populations 
(e.g., Yan et al. 2005; Stark et al. 2007; Eyles et al. 2007).

Nevertheless, these simulations can provide a powerful tool for
understanding the limits in which we can measure structures and
sizes.  The general reason is that these parameters are fairly simple,
and use the galaxy profile and the amount of `clumpy' light in a galaxy
for measurements.  The factor which can differ is the absolute size
of the profile, and the distribution of `clumpy' features, and how large
these are relative to the galaxies themselves.  We addresses these issues 
later in \S 4.2.  However, we discuss
in this section the results of a series of different simulations to determine
the ability to measure galaxy structure in such faint and small galaxies
as within our sample.

\subsubsection{Nearby Galaxy Simulations}

The nearby galaxy simulations we discuss are included in the analysis of
previous papers, including Conselice et al. (2000a), Bershady et al.
(2000), and Conselice (2003).  These simulations consist
of taking nearby, mostly normal, galaxies such as spirals and ellipticals
with some peculiars and irregulars, and rebinning their pixels, and
in some cases adding sky noise, to simulate further distances. The sizes
and structures of these galaxies are then measured the same way they were
before they were simulated to determine how effects of resolution, increased
noise (lower S/N), and higher redshifts, can affect the measured values.

Using a sample of 113 nearby bright galaxies, Conselice et al. (2000a)
determine the resolution limit for which galaxies can have their
asymmetry indices properly measured.  This was done by degrading the
resolution of these 113 galaxies, and then re-measuring the asymmetry
index, and comparing this to the original values.  These simulations and
re-measurements demonstrate that if 0.5 h$_{75}^{-1}$ kpc is resolved 
(or 0.54 kpc in our cosmology), then the asymmetry index can be
measured within 10\%. Furthermore, as Figure~20 of Conselice
et al. (2000a) demonstrates, 0.7-0.9 kpc of structure must be
resolved to utilize this index.  We are imaging our galaxies
at this limit.   However, we note that these
simulations are done for galaxies which are intrinsically larger
than the distant galaxies we are examining in the UDF (e.g., Bouwens
et al. 2004; Ferguson et al. 2004).

In a similar vain, Bershady et al. (2000) simulated 72 nearby galaxies, mostly
spirals and ellipticals, to determine at what limit the concentration index
can be utilised. These simulations are similar to those from Conselice et 
al. (2000a) in that these galaxies were re-binned in their pixels to create
smaller sized systems, and their results are measured in terms of the effective
radius in pixel units.  Bershady et al. (2000) further investigated which
type of concentration index is the most stable to these types of effects,
and therefore the most useful for use at high redshifts, concluding,
as explained in \S 3.2.3 that the index we use here is the most
robust while still providing a broad dynamic range to cover different galaxy 
types.  

However, when galaxies become very small compared to the PSF size, it
becomes more difficult to measure an accurate r$_{20}$ radius from
which the concentration index is measured.  Bershady et al.'s simulations
reduce the half-light radii of galaxies to 0.3-0.7\arcsec, similar to
the largest galaxies within our sample.  Bershady et al. (2000) find
that the half-light radii is extremely stable, and can be measured accurately,
even when sampling just a few pixels.  Furthermore, for galaxies
which have radii measured with 7 pixels, the mean differences in
concentration, relative to the original image, are: 
$\delta C$ = -0.1$^{+0.2}_{-0.6}$, 
increasing to $\delta C = 0.2$ when five pixels are used to sample
the half-light radius, which is only 8\% of the dynamic range given by our
particular choice of concentration index (\S 3.2.3).   About half
of our sample have effective radii smaller than 4 pixels, where the Bershady
et al. (2000) simulations suggest that the measured scatter increases
significantly.  However, we later find that these smallest drop-outs 
display a correlation between half-light radii and concentration (\S 5.4), 
suggesting 
that concentration can be measured even within this potential scatter.

 The Bershady
et al. (2000) simulations are particularly useful for our purposes as
they examine how the concentration index changes as a function of
half-light radii, thus we can directly compare their results with our 
galaxies. Lotz et al. (2004) carry out similar simulations using eight 
galaxies of classes: S0, E, Sab, Sbc, Sc, Sd, and two mergers (the 
Antennae and Arp 220).  They conclude, similar to the findings of 
Conselice et al. (2000a) and Bershady et al. (2000), that the concentration, 
Gini, and M$_{20}$ indices are
reliable to 10\% down to a (S/N) per pixel of $> 2$.  They furthermore
show that concentration and M$_{20}$ can be retrieved to within
15\% down to resolutions of 500 pc, or better and down to 1000 pc for
the asymmetry, Gini and clumpiness indices.   For the reasons above,
and due to limits on measuring structures at small sizes, we place
a restriction of S/N $>10$ and mag $z < 27.5$ on galaxies to be included in
later structural analyses in this paper (Figure~2). These results all suggest
that we are just at the limit, for our smallest and faintest galaxies, 
with the resolving power of ACS and within
our S/N and magnitude limit, to determine accurate parameters for 
our galaxies, within a well defined uncertainty.

\subsubsection{Distant Galaxy Simulations}

The above simulations were however all done using nearby normal galaxies, which
are certainly different from our current $z > 4$ sample, most particularly
within their measured sizes.  As such, we carry out new simulations using
drop-outs themselves, placing B-drops and V-drops to
respective higher redshifts, and measuring how the
structure and sizes of these galaxies change when view in the
redder ACS filters within the UDF.  The general method
for carrying this out is explained in detail in
Conselice (2003).

\vspace{1cm}
\setcounter{table}{0}
\begin{table}
 \caption{Simulation results for various drop-outs placed at higher
redshifts.  These
differences ($\delta$) are such that $\delta = {\rm orig} - {\rm sim}$, that
is the difference between the original images and the simulated
ones.  The values quoted here are the average differences for
the entire simulated sample.  Show are the values for the CAS
parameters, and the total Petrosian radius.}
 \label{tab1}
 \begin{tabular}{@{}ccccc}
  \hline
\hline
Simulation & $\delta$ C & $\delta$ A & $\delta$ S & $\delta$ R \\
\hline
{\tt sim1} & 0.69 & 0.26 & 0.23 & 0.07\arcsec \\
{\tt sim2} & 0.51 & 0.13 & 1.19 & 0.08\arcsec \\
{\tt sim3} & 0.48 & 0.26 & 0.43 & 0.02\arcsec \\
{\tt sim4} & 0.31 & 0.02 & 0.71 & 0.03\arcsec \\
\hline
\end{tabular}
\end{table}

We carry out four different simulations to determine, relative to
our $z = 4$ and $z = 5$ samples, how being more distant would affect our 
measured sizes and CAS values.  The first simulation, which we call 
${\tt sim1}$, is where
we take the B-drops as observed in the V-band and
place them to how they would appear in the $z-$band at
$z \sim 6$ effectively simulating how these galaxies
would appear as $i$-drops observed in the
$z-$band.  The second simulation (${\tt sim2}$) 
takes the V-drops, as seen in the $i-$band and simulates
how these systems would appear in the $z-$band as observed
at $z = 6$. The third simulation (${\tt sim3}$) involves
simulating the B-drops as seen in the i-band into the
$z-$band at $z = 6$.  The final simulation (${\tt sim4}$)
is where we take B-drops as imaged in the V-band and
put them to how they would appear at $z = 5$ in the
$i$-band. 
We carry out these simulations to determine how the CAS values,
fluxes, and sizes change within each of these simulations.
The results of these simulations are shown in Table~1.
Table~1 lists the results of our simulation in terms of
the average differences in the concentration, asymmetry,
clumpiness parameters as well as measurements of the
total Petrosian radius.  The differences are such that
a positive value means the simulated value is smaller
than the original.
We also analyse these simulations in several ways,
including by determining how the various parameters 
change for galaxies of different types, such as
elongated and asymmetric galaxies.  We find very
little difference between the results after dividing
the galaxies into sub-visual types.

The first observation from Table~1 is the large correction
needed to account for differences in the clumpiness index.
We henceforth do not consider this index, partially because
of the difficultly in measuring it, but also due to the
fact that the measurements of this index are difficult
for faint and small galaxies, such as the ones we
examine in this paper (see Conselice et al. 2003a for
a more detailed discussion of this.)  This table also
shows that the concentration index is fairly reliable,
with changes that are generally similar to the measurement
error.  

The asymmetry values are however significantly different
between the two redshifts in these simulations. This can result in 
measured asymmetries
that differ between redshifts, but we note that Table~1
shows that all of these asymmetry changes are such that
asymmetry becomes smaller at higher redshifts, an effect
well known and calibrated using extensive simulations of
nearby galaxies (Conselice 2003; Conselice et al. 2000a).
Therefore, we can and do take the measured asymmetries for
our galaxies as lower limits - the intrinsic asymmetry value
can be higher than the measured value, but not lower, as
redshift effects will only produce a decrease in the
measured asymmetry.

The concentration index also changes slightly in
comparison to the total dynamic range of possible
values.    We also note that the change in the
measure values of the total Petrosian radii,
0.03-0.08\arcsec, are often times much smaller
than the total Petrosian radii which we measure
for our systems.  Furthermore, we utilize these
corrections when discussing the total value of
the asymmetry index from which we make one 
measure of the assembly history for these
systems (\S 4.4).

\subsection{Limits on the CAS parameters}

In summary, we use the CAS parameters in this paper for diagnostics
to determine whether galaxies are in a formation
state, or if they are more likely in a quiescent
mode. However, at these high redshifts, there are limits,
described above in great detail using simulations,
to how much we can use these CAS parameters, and thus how
reliable the results obtained from them are. The problem is that 
although we are able to measure and at times correct for, the 
various effects that can alter
the CAS values, these corrections can be quite
large.  However, as we discuss above, simulations suggest
that we can measure these parameters, at least asymmetry
and concentration, in a reliable way.  Furthermore, our
random errors are accurately representing the scatter
in these values.

Another issue that we must address is that many of our
galaxies are small - some are roughly the size of the PSF or
slightly larger (\S 3.3).  Figure~6 shows the relationship
between the measured size of our objects in arcsec, as a function
of apparent $z-$magnitude.  As we discussed earlier, a fraction of
our systems have half-light radii which are smaller than the
ACS PSF. However, none of our systems have diameters smaller than
the PSF.   Again this implies that we are able to measure accurately
the concentration and effective radii for our samples, which is
also implied by our simulations.

Another way we can determine how the ACS PSF could be affecting
our results is to determine empirically how objects which
are not resolved, namely stars, behave in the various 
diagnostic plots we use throughout this paper.  To carry out
this analysis we use over 3000 stars from the COSMOS
survey.  As an example, the location of these objects in the 
concentration-asymmetry
plane is shown in Figure~7.  The stars in this diagram are coloured
red, and occupy similar concentration values just short of
$C = 1$, with another `branch' nearby at $C = 1.2$.  These stars 
further display a range in asymmetry values.  However, as we discuss
when comparing to the concentration-asymmetry plane, our drop-outs
and most other faint galaxies as labeled on Figure~7,
are not in the same regime as these stars.

\subsection{Internal CAS Error Calibration}

We can test our measured random errors on the asymmetry index, and by
extension, the errors in the other indices, by determining if
the error distribution in the asymmetry parameter is reliable.  We do this by
investigating the error distribution below the asymmetry $A = 0$
limit.  Random errors on the asymmetry index are produced by
background noise that can sometimes dominate the measurement
process.  Conselice et al. (2000a) investigate in detail the
methods for retrieving the error on the asymmetry which we have
applied here.

\begin{figure*}
%\vspace{5.5cm}
 \vbox to 90mm{
\includegraphics[angle=0, width=174mm]{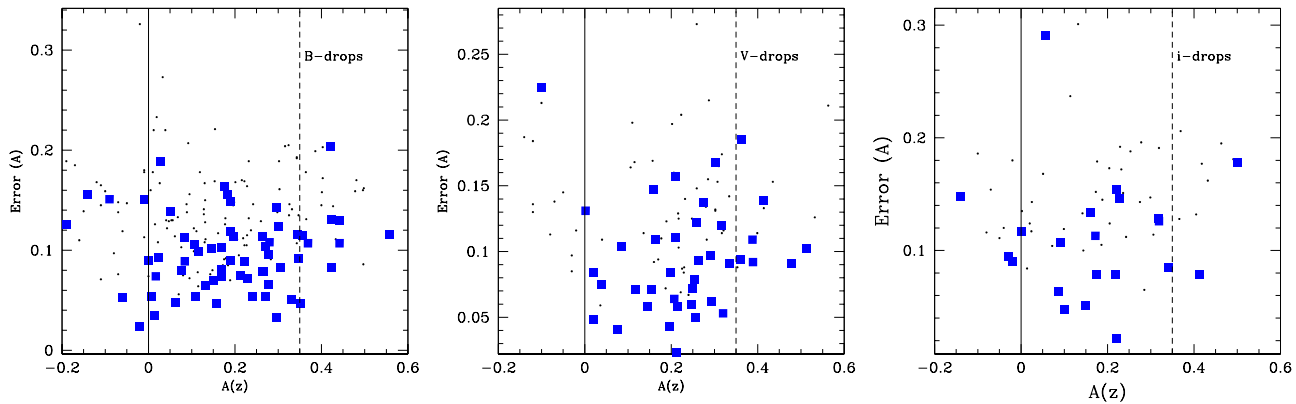}
 \caption{The error distribution in the asymmetry index as a function of
the asymmetry parameter in the $z_{850}$-band.  The dashed line is 
the typical limit used to compute the merger fraction using the rest-frame
 optical, or stellar mass maps, of galaxies.  The solid line is the
$A(z) = 0$ limit.  The red triangles are for those galaxies with
magnitudes $z < 28$, and the smaller black dots are for galaxies
at fainter magnitudes.}
%\vspace{3cm}
} \label{sample-figure}
\end{figure*}

\begin{figure*}
%\vspace{5.5cm}
 \vbox to 80mm{
\includegraphics[angle=0, width=174mm]{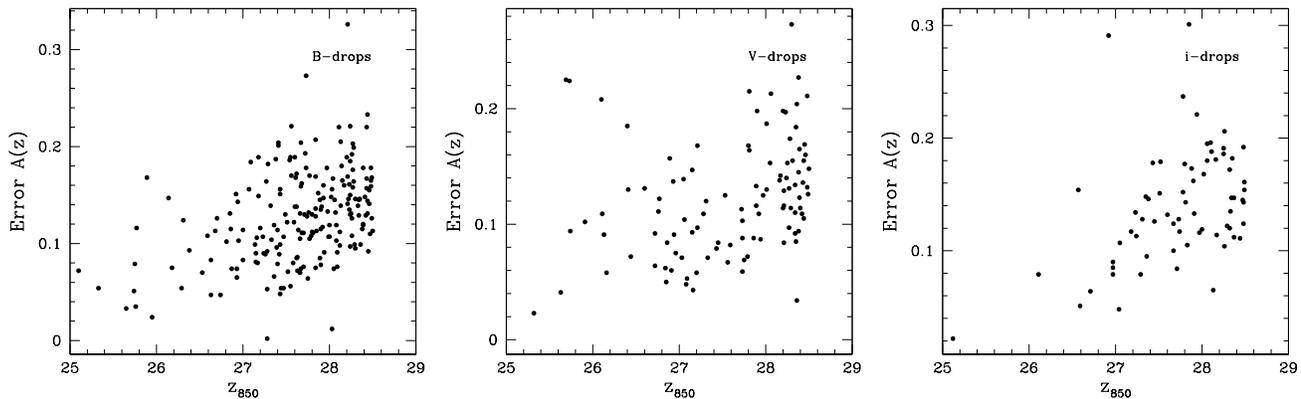}
 \caption{The distribution in the errors in the measurement of
the  asymmetry index as a function of the $z_{850}$
magnitude.  As can be seen, at fainter $z-$band
magnitudes, the error distribution is larger and tends
to have a large spread.}
%\vspace{3cm}
} \label{sample-figure}
\end{figure*}

What we find, as shown in Figure~8, is for each drop-out with $A < 0$, 
there is a correlation between the
asymmetry value and the error on the asymmetry, such that
galaxies with a larger negative asymmetry value have a
corresponding higher error.   For the B-drops, V-drops
and $i-$drops, we find the average asymmetry for galaxies
with $A < 0$ are: $A = -0.14$, $-0.10$, and $-0.09$, respectively.
The average asymmetry errors for these systems are:  $\delta A =$ 0.14, 0.16,
and 0.14,  with the average asymmetry+error: 0.00, 0.06 and 0.05.  As
these random errors match the asymmetry amplitudes at 
$A < 0$, we conclude that the value of these errors
are roughly correct within the high-redshift regime.   We
also do not see any significant trend for the errors to be
lower or smaller for more asymmetric, or less asymmetric
galaxies. This indicates that the high asymmetries are not
due to an Eddington bias, whereby only our largest asymmetries
are found for galaxies with the largest errors.  This issue
is also discussed in detail in Conselice et al. (2003a).   Figure~9
furthermore
shows the distribution of asymmetry error with magnitude, demonstrating,
as expected, that fainter drops have a large error in their
measured asymmetries, which follows for the other parameters as well.

\subsection{Morphological $k-$corrections}

Another important issue that we must account for, to utilise
the optical definition of a CAS merger, is the morphological $k-$correction.
Since we are using the $z$-band imaging for our systems, the
rest-frame light we probe for our drop-outs changes from $\lambda
\sim 1900$ \AA\, to $\lambda \sim 1300$ \AA.  These morphologies
are therefore in the rest-frame UV.  The problem we have with interpreting
structures measured at these rest-frame UV wavelengths is
that we do not know what the rest-frame optical for these
systems, or hardly any galaxies at $z > 1.5$, is compared
to their UV morphologies (cf. Conselice et al. 2005). 
Therefore, there is no simple or
direct way to convert these observed CAS values in the UV
to rest-frame optical ones.

We can however make a best estimate by using observations of
similar systems at $0.5 < z < 1.0$ where we have in the UDF
the rest-frame structures of galaxies from the UV to optical.
Conselice et al. (2008) computed what these morphological
$k-$corrections are for these systems.  For peculiar galaxies
at $z > 0.75$ we find that the morphological $k-$correction 
for the asymmetry parameter is $\delta A/\delta \lambda = -0.83$ 
$\mu$m$^{-1}$.  What this implies is that the asymmetries of the
peculiar galaxies within our sample are too high, from the
morphological $k-$correction, by an amount of $\delta A \sim  -0.29$ to 
$-0.34$ within the redshift ranges we examine.  We use the peculiar
galaxies for this calculation as our visual estimates suggest
these are the correct form to use (\S 5.1).  Furthermore, if we
use other star forming galaxies, such as spirals, we get very
similar results.

As discussed in detail through \S 4 the other major source of systematic
error is produced by the fact that 
these ultra-high redshift galaxies have artificially induced changes in 
their CAS parameters due to the fact that their surface brightness has 
declined significantly, resulting in lower measured CAS values
(\S 4.1).   These effects are luckily changing the measured
asymmetry in opposite directions, such that the rest-frame asymmetry
is close to the observed value.  

The rest-frame optical asymmetry $A_{\rm rest}$
is then the measurement of the asymmetry index $A_{\rm obs}$ plus the change 
due to the morphological $k-$correction $\delta A_{\rm k-corr}$, plus the 
change due to surface brightness dimming $\delta A_{\rm SB-dim}$.    
The net asymmetry can then be written as:

\begin{equation}
A_{\rm rest} = A_{\rm obs}  + \delta A_{\rm k-corr} + \delta A_{\rm SB-dim}.
\end{equation}

\noindent Luckily in our case, the SB-dimming creates apparently
smoother and symmetric systems, such that $\delta A_{\rm SB-dim}$
is positive (e.g., \S 4.1 and Table~1).  The value of 
$\delta A_{\rm k-corr}$ for galaxies observed in the UV is 
negative, and has a value similar to the SB-dimming correction. What
we find is that $\delta A_{\rm SB-dim} + \delta A_{\rm k-corr} \sim 0$,
and therefore opted to not apply any correction to our asymmetry measures
as the SB-dimming correction roughly balances the morphological $k-$correction.
Thus, as best as we can determine, our final asymmetry values are the 
rest-frame optical values, corrected for surface brightness dimming and other
redshift effects.  Further studies will require a longer wavelength,
higher resolution camera than provided by the ACS. WFC3 will provide
longer wavelengths, but not higher resolution, and thus we will likely
have to wait for adaptive optics, or future space missions to carry out
a more detailed analysis able to better limit these biases.

% Normally this problem is dealt with by simulating nearby
%galaxies to how they would appear at higher redshifts and
%then determining how the structures would change due to
%the effects of redshifts. However, this is not likely to
%reveal an accurate systematic in the case of $z > 4$
%galaxies as their sizes and structures are likely
%quite different (e.g., Ferguson et al. 2004; Conselice
%et al. 2008a).  We  test this hypothesis in this
%paper.

\section{Analysis}

\subsection{The Structures of $z > 4$ Galaxies}

Before we discuss in detail the various morphologies, sizes,
and structures of galaxies at $z > 4$, we give a brief
outline of our procedures and our results in this section.  First, 
we present in Figure~10 the morphological breakdown of our sample, as 
defined through visual measurements, at each redshift, as specified
by the drop-out criteria.   What we find is that a large
fraction of our systems are distorted in some way,
as seen by eye.  This suggests that a significant fraction of
the galaxies in our sample are undergoing some type of formation
activity, either through a merger process of some kind, or in some more general
type of assembly.

This fraction is also revealed when applying quantitative methods such
as the CAS merger criteria, and by examining the number of  drop-outs
in pairs.  We stress that each method we use to trace a potential
underlying merger, or assembly event, has significant uncertainties 
associated with it. However, as we quantify throughout, these
different methods all suggest similar results - that roughly
30\%-50\% of drop-out galaxies are likely in some kind of
dynamically active phase, perhaps produced by the merger
process.  It is important to note the corollary of this, which
is that 50-70\% of the drop-outs appear symmetrical, without any
significant sub-structure.

\begin{figure}
%\vspace{5.5cm}
 \vbox to 140mm{
\includegraphics[angle=0, width=90mm]{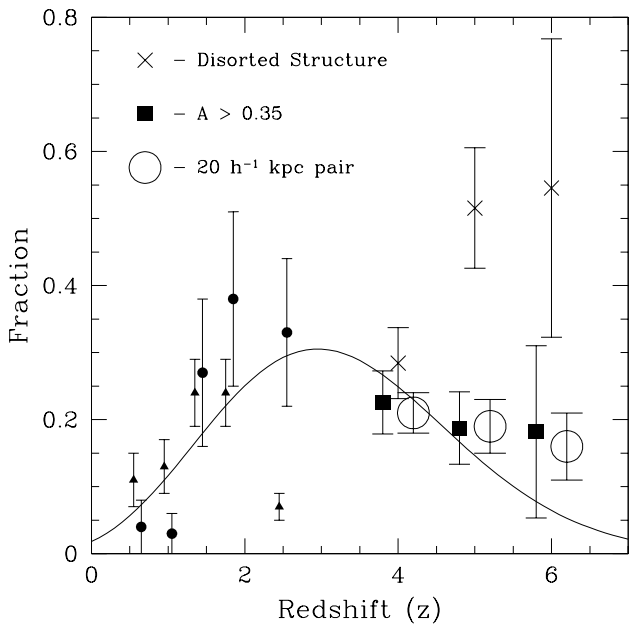}
 \caption{The fraction of galaxies in a likely dynamical
assembly state, as
a function of redshift using different selection methods.
The solid dots at $z < 3$ are points taken
from the combined UDF+HDF CAS analysis in Conselice 
et al. (2008) using a M$_{*} > 10^{10}$ \solm
selected systems, while the triangles at $z < 3$
are for galaxies with masses $10^{9}$ \solm $<$ M$_{*} < 10^{10}$ \solm.  
These are mergers selected using the criteria: $A > 0.35$ and $A > S$.   
The crosses are those LBGs which have a distorted structure as 
judged visually, the solid boxes are systems which are consistent with 
merging within the CAS criteria of $A > 0.35$ and $A > S$, 
while the open circles at $z > 4$ are the fraction of galaxies
which are within pairs.  The solid line shows the
best-fit Press-Schechter based form (roughly a exponential/power-law
combination fit) for how the merger
fraction evolves with time using the CAS criteria
to locate mergers at $z > 4$. }
%\vspace{3cm}
} \label{sample-figure}
\end{figure}

Table~2 and Figure~10 presents a summary of our findings concerning the
distribution of galaxies in various inferred merger and structural
states at $z > 4$.  The fraction of galaxies which appear
by visual inspection in the $z-$band
to have a distorted or merging structure (f$_{\rm merger}$)
varies between $\sim 30$-60\% for the drop-outs.   These
are the fraction of systems which we classify by
eye as either class (iii) or class (iv) in the visual
typing (\S 3.1).  The remainder of the systems, or
the fraction ($1 -$ f$_{\rm merger}$), have a smooth
structure, in classes (i) or (ii).  These normal
galaxies, which make up the bulk of the population
seen by eye, are discussed in terms of their
formation modes and history in \S 5.4.  This is largely
the only use that we will make of these visual morphologies,
although we do discuss them in context of other
quantitative results in the discussion.

\begin{figure*}
%\vspace{5.5cm}
 \vbox to 80mm{
\includegraphics[angle=0, width=174mm]{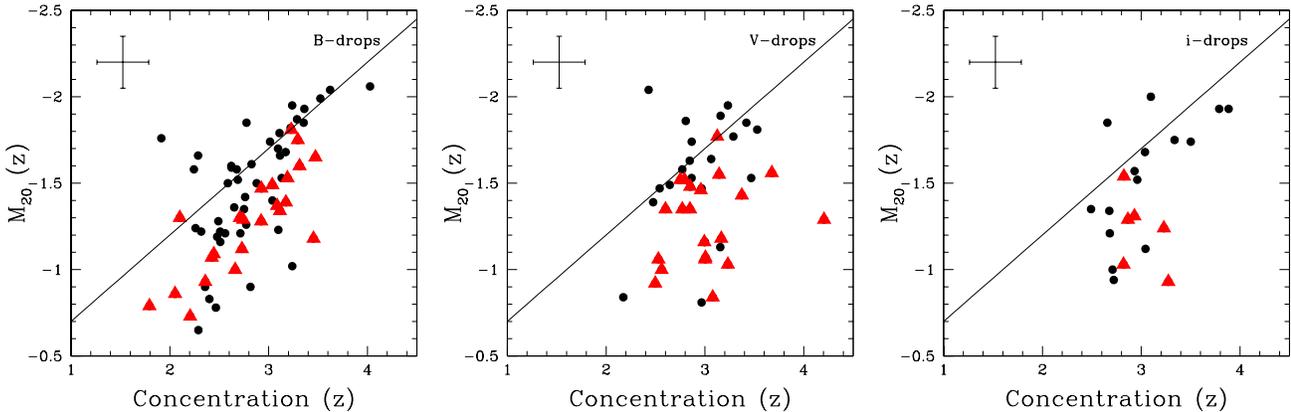}
 \caption{The correlation between the concentration index ($C$) and
the moment ratio M$_{20}$ in the observed $z-$band.  The solid round
points are for galaxies identified visually in the $z-$band image
as normal and symmetrical systems, without any obvious peculiar
structure, while the red triangles are for those systems which
are visually identified as having a peculiarity.  The solid
line show the relationship between $C$ and $M_{20}$ for the most
concentrated systems with $C > 3$, and which have no peculiarity
in their structure. The error bars represent the average error in
our measured $C$ and M$_{20}$ values.}
%\vspace{3cm}
} \label{sample-figure}
\end{figure*}

\vspace{1cm}
\setcounter{table}{1}
\begin{table}
 \caption{Measured merger fractions and pair fractions for drop-outs.  
The value of f$_{\rm merger}$ is the merger fraction determined by
visual estimates, f$_{\rm asym}$ is the fraction of asymmetric
galaxies, and f$_{\rm pair}$ is the ratio of the number of pairs
to the total galaxy population (see text). }
 \label{tab1}
 \begin{tabular}{@{}cccc}
  \hline
\hline
Drop-out & f$_{\rm merger}$ & f$_{\rm asym}$ & f$_{\rm pair}$ \\
\hline
B-drop & 0.28$\pm$0.05 & 0.23$\pm$0.05 & 0.21$\pm$0.03 \\
V-drop & 0.52$\pm$0.09 & 0.19$\pm$0.05 & 0.19$\pm$0.04 \\
i-drop & 0.55$\pm$0.23 & 0.19$\pm$0.13 & 0.16$\pm$0.05 \\
\hline
\end{tabular}
\end{table}

We can get some sense of the structural properties
of our sample of drop-outs by examining the relationship between
the concentration index ($C$) and the moment ratio parameter,
M$_{20}$.  Both of these parameters measure how much light
is concentrated in galaxies.  The $C$ parameter differs from
M$_{20}$ in that the $C$ parameters measures the concentration
with regards to the centre of a galaxy, and M$_{20}$ gives more
spatial information on where light is distributed, and is
more sensitive to outer light than the concentration.  A galaxy
with a high $C$ value, e.g., $C > 3$, should have a low
M$_{20}$ index with values $M_{20} < -1.5$.

In general, we find that galaxies with a high $C$ index are 
normal, without visual evidence for peculiarities (e.g.,
Figure~11), and these galaxies also have the lowest
$M_{20}$ values.  If we extend the rough correlation between
$C$ and M$_{20}$ for the high $C$ objects, we do not find
objects which have the same (low) M$_{20}$ values per given
value of the concentration index.  Indeed, what we find is
that the values of M$_{20}$ get larger per concentration
value, at lower concentrations.    What this means is that
for non-concentrated galaxies, the distribution and brightness
of the brightest 20\% of pixels grows larger at a lower
concentration.  This is one indication that these systems
are not in a relaxed state. In fact, the galaxies
which deviate most in M$_{20}$ are the galaxies that are
visually identified as merging systems, but not uniquely
so (Figure~11).

\subsubsection{Peculiar and Asymmetric Systems}

We derive the unusual or non-symmetric fraction of 
galaxies within our sample in three different ways. This 
includes
investigating how many peculiar galaxies, as measured
by eye, we have
in our sample at each redshift, how many galaxies
are quantitatively asymmetric and have uneven light
distributions suggestive of mergers/assembly, as well
as a new technique to find pairs using the 
Lyman-break methodology to determine whether two
galaxies projected on the sky are potentially
merging.  Each of these methods has
systematic errors which we address
quantitatively.

Perhaps the most straightforward method for determining
the merger fraction for these Lyman-break drop-out
galaxies at $z > 4$ is to determine how the fraction
of distorted and peculiar galaxies changes with
redshift.  First we examine the fraction of drop-outs which
appear visually peculiar.  We show this evolution in 
Figure~10 (as crosses), where we find that the fraction of
galaxies in our drop-out sample, with mag $z < 27.5$ and
S/N $> 10$, which have a peculiar structure,
possibly indicative of mergers or recently assembly, ranges 
from 0.28$\pm$0.05 for the B-drops to 0.52$\pm$0.09 for the V-drops, and
0.55$\pm$0.23 for the $i-$drops.   What is perhaps surprising
about these numbers is that roughly half of all the drop-outs
appear to be symmetrical and round systems that are unlikely
to have gone through a recent dynamical assembly episode.

\begin{figure*}
%\vspace{5.5cm}
 \vbox to 80mm{
\includegraphics[angle=0, width=174mm]{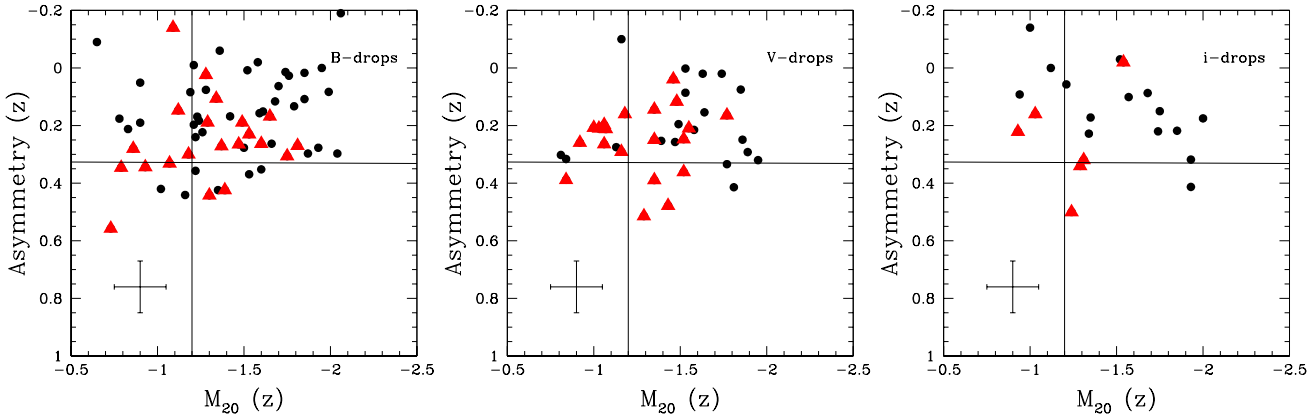}
 \caption{Similar to Figure~11, but showing the relation
in the observed $z-$band between the asymmetry of our
sample and the M$_{20}$ index. In general we find that
galaxies which are more symmetric, have lower (more
negative) M$_{20}$ values.  As in previous figures, the
red triangles show the location of galaxies identified
as peculiar.  The solid horizontal line shows the $A > 0.35$
limit for finding 'mergers', while the vertical solid
line shows the limit for finding galaxies merging with
the Gini/M$_{20}$ system, with the criteria M$_{20} > -1.2$.
The error bars represent the average error in
our measured $A$ and M$_{20}$ values.}
%\vspace{3cm}
} \label{sample-figure}
\end{figure*}

\begin{figure*}
%\vspace{5.5cm}
 \vbox to 80mm{
\includegraphics[angle=0, width=174mm]{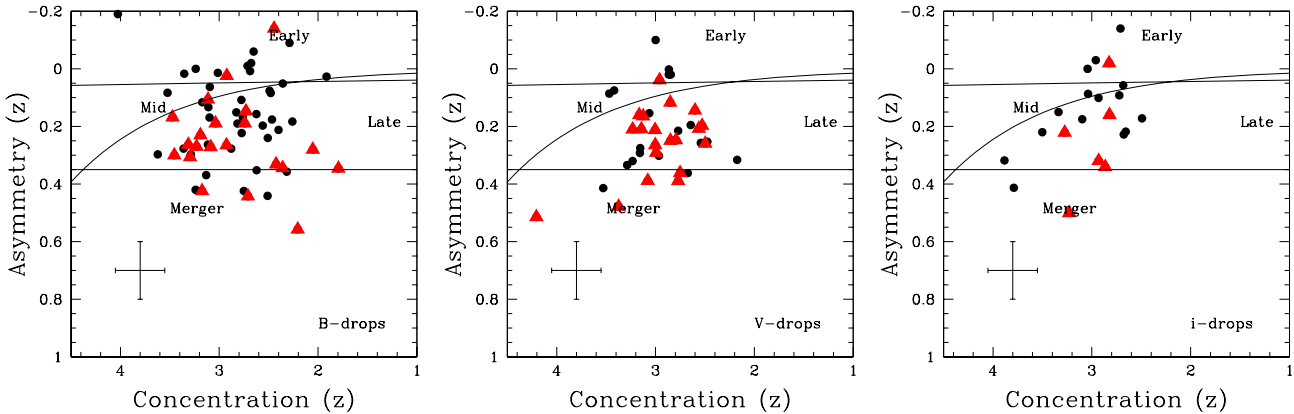}
 \caption{The concentration-asymmetry diagram for galaxies
at $z > 4$.  Each of the three panels shows the C-A diagram
for those systems which are B-drops, V-drops and $i$-drops.
The lines and labels denote the various areas of this space
where galaxies of different types are found within the
nearby universe.  As in previous figures, the triangles
represent those systems which by eye appear to be distorted,
or in some type of merger phase, and the dots are those
systems that look more normal.  The error bars represent the average 
error in our measured $C$ and $A$ values.}
%\vspace{3cm}
} \label{sample-figure}
\end{figure*}

We can also see this diversity in the structures of our drop-out
samples by investigating where they fall in structural diagrams.
First, as already discussed, Figure~11 shows that galaxies which
are chosen by eye as peculiar have a high M$_{20}$ index at lower
concentrations, an indication that systems that appear as peculiar
have a significant fraction of their light in their outer parts, rather
than concentrated towards the centre.   Similar trends can be
seen when we examine the asymmetry index with the M$_{20}$
index (Figure~12).  The M$_{20}$ index has been used previously
as a sole indicator for finding mergers in rest-frame UV
imaging (e.g., Lotz et al. 2006), such that systems which have
a high M$_{20}$ index are more likely to be systems undergoing some
form of assembly or merging.

Figure~12 shows that for the B-drops and the V-drops there is a 
rough relation
between asymmetry and M$_{20}$, such that galaxies which are more
asymmetric have higher M$_{20}$ values.  We also find
that those systems which appear visually as peculiar (labeled as
triangles) are more likely than non-peculiar and smooth systems
(\S 3.1; small dots) to have
a high asymmetry and/or a low M$_{20}$ value.   In fact the
only systems with low-asymmetry and low M$_{20}$ values are
the smooth normally appearing galaxies.  This is a 
verification that our methods for identifying structurally
smooth galaxies with the CAS and M$_{20}$ parameters works.

Figure~13 shows the location of our sample within 
concentration-asymmetry diagrams.  Similar to Figure~12, we
find a mixture of structures, as measured quantitatively, for
our sample, with a range of light concentrations and asymmetry
distributions.  We also label on Figure~13 the location of nearby
galaxy types. Although we are not arguing that these drop-outs
are similar in anyway to nearby galaxies, it does show that
the measured CAS values span the range of the values found for
nearby galaxies.

We use the simple merger fraction criteria from the CAS
method of $A > 0.35$ to calculate the merger fractions for our
drop-out samples.  Note that the CAS values we plot on Figures~11-13
are the observed values, and we must use the corrected CAS values
for k-corrections and redshift (\S 4.4) to measure a merger fraction in
a comparable way to the measures at lower redshifts (e.g., Conselice
et al. 2008).   What we find is that the inferred merger fractions using
these rest-frame optical and calibrated asymmetries values 
are: 0.23$\pm$0.05 (B-drops), 0.19$\pm$0.05 (V-drops)
and 0.19$\pm$0.13 (i-drops).  

One aspect that Figures~12-13 shows is that there are some
galaxies which are identified as a merger by visual
estimates, but which do not have very high asymmetries.
This is a well known effect, and has been  documented
and discussed in Conselice (2003a), Conselice et al. (2008);
Lotz et al. (2008),
and other papers.  For nearby ongoing major mergers, such
as ULIRGs, only roughly half of these systems have a large
asymmetry, such that $A > 0.35$ (Conselice 2003).  This is due to 
the fact that within the merger process, which can last for over
a Gyr, only during a fraction of this time will the merger
be identified as having a high asymmetry (e.g., Conselice 2006).  
N-body models show that the time for a galaxy to have
a high asymmetry is roughly a factor of 2-3 times shorter
than the entire merger process (Conselice 2006; Lotz
et al. 2008b).  During other times,
the merging galaxy will fall into non-merger
regions of the CAS space.

We can however test the likelihood, based on simple arguments,
that these two determinations are measuring part of the
same population.  As Figure~13 shows, only a fraction
of the systems with $A > 0.35$, or at least galaxies with a high
asymmetry, are found to have a structure which is classified
as a peculiar or merger through our visual estimates. This
is also reflected in the higher fraction of galaxies which
look peculiar compared with those that are highly asymmetric
(Figure~10).

\subsubsection{Dynamical Time-Scales}

We can use the fact that some of our galaxies are peculiar, and
the assumption that these structures are produced through
either mergers or some type of assembly, to calculate the likely
dynamical time-scales for these galaxies.
If we take as our hypothesis that these distorted structures
are tracing changes in the underlying potential due to a
merger, then we calculate through basic arguments
the amount of time the galaxy will appear distorted
given its size and internal velocities. 

A very important question within this analysis is how long
a galaxy which has recently undergone a merger will appear
distorted.  The dynamical time ($\tau_{\rm dyn}$), or crossing time, of
a galaxy can be written as,
%= \frac{R}{\sigma_{\rm int}}

\begin{equation}
\tau_{\rm dyn} = \left(\frac{\rm R^{3}}{\rm G\,M_{dyn}}\right)^{1/2} \sim 10^{7} {\rm years} \left(\frac{R}{1\, {\rm kpc}}\right) \left(\frac{100\, {\rm km s}^{-1}}{\sigma_{\rm int}}\right),
\end{equation}

\noindent where ${\rm R}$ is the radius of the
galaxy, M$_{\rm dyn}$ is the total mass, and $\sigma$ is the internal velocity
dispersion.   This equation was used previously by e.g., Hathi et al.
(2008) to measure the time a symmetrical galaxy must be dynamically
quiescent.  For our systems, we obtain dynamical time-scales 
of $< 10^{8}$ years.  
However, this time-scale is unlikely to represent the time for
a galaxy to become relaxed and produce a normal and smooth
galaxy profile.    More generally, we are interested in how
long a galaxy appears morphological peculiar after a merger
event.  We can use the two-body relaxation
time expanded to larger systems, in our case a galaxy.  However, the
two-body relaxation time is very long, as the crossing
times of a galaxy are too long for star-star encounters to
produce effective relaxation. 

As described in Lynden-Bell (1967), the time-scale for a merging galaxy
to become relaxed is determined by the rapidly changing potential
energy within the evolving system.  This `violent relaxation' is
a complex process which is difficult to characterise through an
analytical approximation, and the best way to approach measuring
morphological time-scales is through N-body simulations of
the merger process (e.g., Mihos 1995; Conselice 2006; Lotz
et al. 2008b).  Mihos (1995) attempted to determined the
amount of time that two disk galaxies would be visible  
as merging systems when observed with the Hubble Space
Telescope's Wide Field Camera-2 at $z = 1$ and $z = 0.4$
within a 10$^{4}$ second exposure (note that the UDF exposures
at about 100 times this) using the F785LP filter. 
 Mihos (1995) conclude that merger
features can be seen for around 350 Myr at $z = 0.4$, but
for a shorter period of time at $z = 1$.  

However, these simulations were done using the less efficient and lower
resolution WFPC2 camera, and using a shorter exposure times than our 
ACS images.  The question of how long a
merger can be identified through CAS and visually
was addressed in Conselice (2006), and Lotz et al.
(2008b) who both found that the visual mergers last for
on order 1 Gyr, and that the CAS method would find mergers
for 0.5 Gyr.  Based on the CAS method we can conclude that the 
time-scale for peculiars to have formed from mergers
would be $<$ 0.5 Gyr in the past, as they are asymmetric.  
Likewise, observational conditions assumed in Mihos (1995)'s simulations
for $z = 0.4$ are similar to the conditions (e.g., S/N per total
observation) for galaxies observed at $z = 5$.  Thus, we conclude
that the visual estimates for finding a merger for our systems
are no shorter than 0.4-0.5 Gyr.  

%This time-scale, which
%we call the relaxation time, is for a galaxy made up of an
%unrelastic number of star $N$ with the same mass:

%\begin{equation}
%\tau_{\rm relax} = \frac{\sigma^{3}}{4\piG^{2}m_{\rm star}^{2}n ln \Lambda} \approx \frac{0.06N}{\rm {ln} (0.15n)} \times \tau_{\rm dyn},
%\end{equation}

%\noindent where $\sigma$ is the velocity of a star, $m_{\rm star}$ is the
%mass of the stars, $n$ is the number density of stars, $\Lambda$ is the
%ratio of maximum to minimum impact parameters, and $\Tau_{\rm dyn}$ is
%defined above.

This time-scale tells us not only how long a galaxy would be seen
as asymmetric, but also reveals, for a smooth system, the minimum amount of 
time since
the last major merger or assembly episode. We also know from simulations 
that the CAS method will only
pick out a merger within a given amount of time, roughly
1/2 to 1/3 of the entire merger process (Conselice 2006). If this is the
case then we would expect the ratio of the visual merger
fraction (f$_{\rm merger}$) to the ratio of the CAS
merger fraction (f$_{\rm asym}$) to be between
2-3. We calculate ratios of 1.2, 2.7 and 2.9, in rough agreement 
with expected values, if asymmetries and peculiar structures
are tracing a larger potential. It is important to
note that the galaxies we examine are likely dominated
throughout their structures by star formation, and
therefore the UV images of these systems reveals
their structure, unlike at lower redshifts
where galaxies contain a mixture of young and more
evolved stellar populations (e.g., Windhorst et al. 2002;
Conselice et al. 2000c; Taylor-Mager et al. 2007).

\subsection{Pair Fraction from Lyman-Break Galaxy Pairs}

It has often been commented on in the high$-z$ galaxy literature 
that drop-outs seen in the GOODS and the UDF fields are found
in pairs - that is two drop-outs appear to be near each other
in the sky.  Unlike the case for general field galaxy populations,
it is relatively straightforward to determine whether two
galaxies, which are drop-outs, are likely true physical
pairs as opposed to simply chance superpositions due to the
nature of the Lyman-break.  Example of these Lyman-break
galaxies in pairs are shown in Figure~3.

We measure the Lyman-Break galaxy pair fraction by utilising the 
feature of the Lyman-break, which limits
the range on the redshifts of galaxies, to isolate galaxies
at similar narrow redshifts ranges.  This allows us to determine,
with a high certainty, the merger fraction for these systems, as it
automatically removes galaxies at very different lower and
higher redshifts.  While there is still a probability that these
galaxies are chance superpositions within the break redshift range,
in practice, the surface density is low enough that this correction
is fairly minor.  

We define a galaxy Lyman-break pair by those which are separated
by 20 h$^{-1}$ kpc, or less.   We do not use a magnitude limit to
select our pairs.  We do this so as to not bias the measured values
due to unknown k-corrections for these galaxies, which are often not
resolved into separate systems within {\em Spitzer} imaging.   Therefore
our values are not proper measured merger fractions within the normal
1.5 B-mag range.   We define the number of galaxies
which are within our separation using each of our Lyman-break
selections as N$_{\rm pair}$, and the merger fraction is thus,

\begin{equation}
f_{\rm pair} = \frac{N_{\rm pair}}{N_{\rm tot}} - {\rm cor},
\end{equation}  

\noindent where 'cor' is a correction for the background given by the
ratio of the average number of galaxies found within an aperture of
20 h$^{-1}$ kpc in radius, but placed randomly throughout the UDF image.
We also take into account the fact that some galaxies near the edges of
the frame cannot have a properly measured pair fraction due
to the limited survey area (e.g., Patton et al. 2000). 

We make a correction for the limited total area of the survey by 
considering the ratio of the area in which a pair
is identified ($A_{\rm pair}$), and the total area covered
in the survey ($A_{\rm survey}$).   Considering the situation
where the observed pair fraction ($f_{\rm pair,obs}$) is known, 
then the true pair fraction ($f_{\rm pair,real}$) is given by:
      
\begin{equation}
f_{\rm pair,real} = \frac{1}{1-A_{\rm pair}/A_{\rm survey}} \left(f_{\rm pair,obs} - \frac{A_{\rm pair}}{A_{\rm survey}}\right).
\end{equation}

\noindent After correcting for the edge of the field, we find that
the pair fraction for the B-drops, V-drops and $i$-drops are:
0.21$\pm$0.03, 0.19$\pm$0.04 and 0.16$\pm$0.05, respectively.
These values are all within 1 $\sigma$ of the CAS values, therefore
two independent methodologies are able to retrieve the same
value for the merger history at $z \sim 4 - 6$.  Examples of
these physical pairs, where both galaxies are a Lyman-break drop-out
at the same redshift, are shown in Figure~3.   Note that if a galaxy is
in a pair, we do not rule out that it can also be
counted as a merger based on its structural parameters.
 
\subsection{The Inferred Evolution of Galaxy Assembly}

\subsubsection{Outline}

We use the results of the previous sections, that is the
peculiar galaxy fraction (\S 5.1) and 
the incidence of galaxies in pairs (\S 5.2) to determine the 
assembly and possible merger state of LBGs at $z > 4$.
As discussed in \S 5.1, the fraction of galaxies which
appear peculiar, by eye, varies between $\sim 0.3-0.6$ 
for the drop-outs.  The fraction of galaxies which are mergers,
based on the CAS criteria, varies between $\sim 0.2-0.25$ within
the same redshifts.  The result of this is that the implied
merger fractions for both the visual identifications, and
for the CAS method are similar, although the visual method
does find a slightly higher fraction (\S 5.1; Figure~10). We can use these
merger fraction estimates to calculate the merger fraction evolution for
our systems by comparing directly to the merger fraction
measurements at $z < 3$ taken from Paper I.

Since we have previously measured our merger fractions as
a function of stellar mass, we need to have some understanding
of the stellar masses for our $z > 4$ galaxies.  The spectral
energy distributions for these drop-outs have been studied
in detail by Yan et al. (2005, 2006), Stark et al. (2007) and 
Eyles et al. (2007) who find that brighter UDF drop-outs, typically
those with Spitzer detections have typical
stellar masses of  $\sim 10^{10}$ \solm (Yan et al. 2005). 
However, most of the high redshift drop-outs, particularly 
at $z \sim 6$, are not detected with Spitzer, and have
stellar masses lower than  M$_{*} = 10^{10}$ \solm (e.g.,
Stark et al. 2009).  In fact, the results of 
Stark et al. (2009) suggests that at the faintest
bins we consider, the stellar masses of our objects
range from $\sim 10^{9-10}$ \solm.

No full analysis of the stellar masses of
drop-outs in the UDF have been published, although
we can use the above arguments to suggest which stellar
mass of galaxies our drop-outs should be compared with.
Figure~10 shows the merger fraction for our drop-outs
compared with two galaxy samples at $z < 3$: those with
stellar masses M$_{*} > 10^{10}$ \solm and
$10^{9}$ \solm $<$ M$_{*} < 10^{10}$ \solm as discussed
in Conselice et al. (2008).  What is found is that
the merger fractions for these two mass ranges are
very similar except for the highest redshift
point at $z \sim 2.5$, where the lower stellar mass
sample merger fraction begins to decline. While our 
drop-out samples may
be dominated by galaxies with stellar masses
M$_{*} < 10^{10}$ \solm, the higher merger fraction,
and the intense star formation rates of these
drop-outs (e.g., Stark et al. 2009), which rapidly
increases their stellar mass, suggests
that these systems are better compared to the higher
stellar mass limit at $z < 3$.

We hence compare
our measured morphological and structural merger fractions
for our $z > 4$ drop-outs with galaxies of stellar masses
M$_{*}$ $> 10^{10}$ \solm  at $z < 3$ as taken from 
Conselice et al. (2008),  and as plotted on Figure~10.  
Note that when we do use a lower
stellar mass limit to compare with, such as 
M$_{*}$ $> 10^{9}$ \solm, we find very similar
results to that presented here.  

\subsubsection{The Evolution of Mergers}

We use the results from \S 5.1.1 and \S 5.2 to derive
the evolution of the merger fraction up to $z = 6$. As far
as we are aware this is the first attempt at measuring the
galaxy merger history back to these early epochs.   Earlier
papers have investigated this history at $z < 3$ (e.g.,
Conselice et al. 2003a; Conselice 2006; Conselice
et al. 2008; Bluck et al. 2009).
  
Before we examine possible merger histories it is worth
reviewing the caveats and assumptions we have made, as
well as the factors which show that we are indeed able
to make a reliable measurement.  First, because of the
distances to these galaxies, and the unknown state of their
gas and structure, we cannot make any firm conclusions
regarding their merger state.  We can infer the that 
structural parameters are measuring
some type of assembly, but whether this is major mergers,
minor mergers, or some type of gas accretion event is
unknowable within our data.  Although high asymmetries are
thought to signify major mergers at at least $z < 3$, this 
may not be the case at $z > 4$ where conditions are
quite different.  Therefore, what we claim to be a measured
merger fraction or evolution can be interpreted as some type
of assembly event that has not cooled dynamically. 

However, the fact that the
CAS method for finding mergers is within $< 1\, \sigma$ of
the galaxy pair fraction, similar to the situation at
$z \sim 0$ (De Propris et al. 2007), is a strong
indication that we are potentially probing correctly the merger
fraction through both methods. This is also reinforced
by the roughly factor of two higher fraction of
distorted galaxies than systems which are asymmetric, again
similar to galaxies at $z < 3$ (e.g., Conselice et al. 2005;
Conselice et al. 2008).  With this caveat we are now able to
trace what is the inferred evolution of this assembly.

The traditional method for parameterising the evolution of
the merger fraction is to use a power-law fitting
formula of the form:

\begin{equation}
f_{\rm m}(z) = f_{0} \times (1+z)^{m}
\end{equation}

\noindent where $f_{\rm m}(z)$ is the merger fraction at a given
redshift, $f_{0}$ is the merger fraction at $z = 0$, and
$m$ is the power-law index for quantifying the merger
fraction evolution.  However, as can be seen in Figure~10 the
merger fraction evolution for the pairs levels off, and there
is not as great an increase at higher redshifts.  This levelling off
has been seen for the general galaxy population at $z < 3$
by Conselice (2006), Conselice et al. (2008) and Ryan
et al. (2008), and can be fit by the Press-Schechter
inspired merger fraction form (Carlberg 1990):

\begin{equation}
f_{\rm m} = \alpha (1+z)^{m} \times {\rm exp}(\beta(1+z)^{2}),
\end{equation}

\noindent where the $z = 0$ merger fraction is given by
$f_{\rm m}(0) = \alpha \times$ exp($\beta$). We use this
fitting function for our merger fraction, utilising the
results of both the pair and the CAS method for
determining the merger fraction history up to $z \sim 6$,
the result of this is shown as the solid line on Figure~10.
We also fit a combined power-law/exponential, as is done in
some previous work (e.g., paper I), of the form:

\begin{equation}
f_{\rm m} = \alpha (1+z)^{m} \times {\rm exp}(\beta(1+z)),
\end{equation}

\noindent and find that it gives a similar fit, based on
the $\chi^{2}$.   However,  these forms do not completely
fit the merger history at $0 < z < 6$, and in fact,
there is no simple way to parameterise the currently
known merger fraction at $0 < z < 6$ selected at a constant stellar
mass.  We explore several fitting
routines, and find that the best-fit two parameter model
is an exponential/power-law of the form:

\begin{equation}
f_{\rm m} = \alpha (1+z)^{3} \frac{1}{{\rm exp}(\beta \times z)},
\end{equation}

\noindent which is designed such that $f_0 = \alpha$, and is only a 
two parameter
parameterisation, as opposed to the three parameter exponential/power-law
discussed above. This form also fits the data as well as the three parameter
models above.  In any case, we have found that fitting
with the exponential/power-law and the Carlberg (1990) version of
the exponential/power-law, gives an exponent on the power-law
portion of $m = 3$.  We find however, that $m = 3-5$ give almost
as good of a fit, as $m = 3$.  

One possible reason why no simple parameterisation is well fit by
the data is that the galaxies that make up the merger fraction at
the highest redshifts evolve to become more massive than the comparison
sample at lower
redshifts.  Because we are using a constant stellar mass limit for
determining the merger fraction, we are comparing galaxies with
stellar masses M$_{*} > 10^{9-10}$ \solm at all redshifts. However the
galaxies at $z > 4$ we examine are possibly, due to future mergers
and star formation, to be among the most massive at lower redshifts, such
that a better comparison is possible when using a higher stellar mass
limit at the lower redshifts.  Ideally we want to trace the same
galaxies and how their merger histories evolve through time.
When better data becomes available
it will be possible to measure these parameters more accurately,
and eventually trace the same galaxies through time based on their
star formation and merger histories.

%black body - (1/(exp(c2*(1+x))-1))*c1*(1+x)**3

%black bodymod - (1/(exp(c2*x-1)+1))*c1*(1+x)**3
%exp(c3*(1+x))*c1*(1+x)**c2
%exp(c3*(1+x))*c1*(1+x)**c2
%exp(c3*(1+x)**2)*c1*(1+x)**c2

% c1*((1+x)**3)*exp(c3*(1+x)**2)
% c1*((1+x)**c2)*exp(c3*(1+x)**2)
% c1*((1+x)**c2)*exp(c3*(1+x))

\subsubsection{Galaxy Merger Rates}

We furthermore compute the galaxy merger rate per galaxy, or
the value $\Gamma = \tau_{\rm m}/f_{\rm gm}$ (see paper III, Conselice
et al. 2009 for a detailed explanation of this.) Where we 
convert the merger fraction ($f_{\rm m}$) into the galaxy merger
fraction (Conselice 2006), through the equation:

\begin{equation}
f_{\rm gm} = \frac{2 \times f_{\rm m}}{1 + f_{\rm m}}
\end{equation}

\noindent The value of $\Gamma$, defined in this way, measures how long an 
average galaxy will evolve passively before undergoing a merger.  The inverse
of $\Gamma$ integrated over time gives the average number of
mergers a galaxy undergoes between two redshifts.

\begin{figure}
%\vspace{5.5cm}
 \vbox to 140mm{
\includegraphics[angle=0, width=85mm]{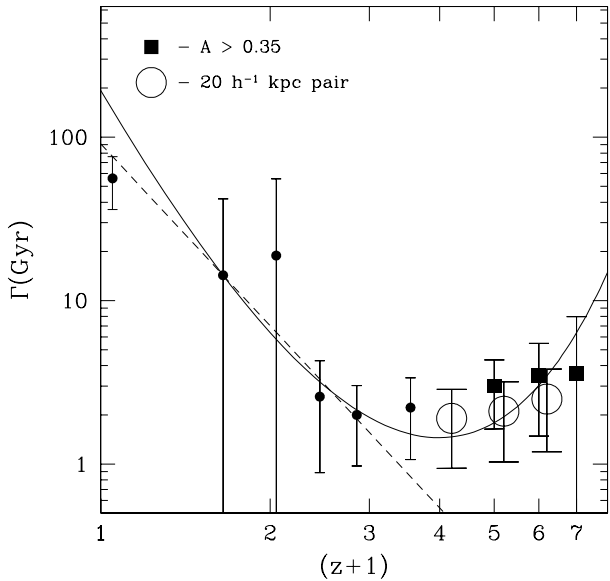}
 \caption{The evolution of $\Gamma$, the average time between
mergers for galaxies with M$_{*} > 10^{10}$ \solm, as a function of
redshift.  The values shown
are for those selected with $A > 0.35$, and those which are within
20 h$^{-1}$ kpc pairs.   The points at $z < 3$ are taken from a combined
UDF+HDF sample from Conselice et al. (2008), using the
CAS methodology.  The solid line
shows the best fit power-law+exponential parameterisation of the
evolution for $\Gamma$, while the dashed line shows the best fit
power-law, which vastly under-predicts the value of $\Gamma$ for
systems at $z > 3$.}
%\vspace{3cm}
} \label{sample-figure}
\end{figure}

Figure~14 shows the evolution of $\Gamma$ within our sample. An important
issue when calculating $\Gamma$ is the time-scale in which the
CAS system is sensitive to the merger process, which we denote as
$\tau_{\rm m}$. We utilise several time-scales, including the 
time-scale calculated in Conselice (2006) ($\tau_{\rm m}$ = 0.34 Gyr)
and the average time-scale for CAS mergers published in Lotz et al. 
(2008b) ($\tau_{m} = 1.0\pm$0.2 Gyr) to calculate the total
number of mergers a galaxy at $0 < z < 6$ undergoes.    In fact,
between two redshifts $z_1$ and $z_2$ the total number of mergers a 
galaxy will undergo ($N_{\rm merg}$) is given by,

\begin{equation}
N_{\rm merg} = \int^{t_2}_{t_1} \Gamma^{-1} dt = \int^{z_2}_{z_1} \Gamma^{-1} \frac{t_{H}}{1+z} \frac{dz}{E(z)},
\end{equation}

\noindent where $t_{H}$ is the Hubble time, and $E(z) = [\Omega_{\rm M}(1+z)^{3} + \Omega_{k}(1+z)^{2} + \Omega_{\lambda}]^{-1/2}$ = $H(z)^{-1}$. 
The result of this calculation using $\Gamma$ is shown in Figure~15, with
an additional time-scale of $\tau_{\rm m} = 0.5$ Gyr shown, including
0.35 Gyr and 1.0 Gyr. 
Using equation (13) we compute that from $z = 6$ to
$z = 0$, the number of mergers a galaxy with M$_{*} > 10^{9-10}$ \solm
undergoes
depends strongly on the adopted value of the CAS merger time-scale
($\tau_{\rm m}$), as shown in Figure~15.  The range in the total number of 
mergers is $N_{\rm merg}$
= 2.5 to 7, depending on the time-scale used.  In fact, by integrating the
individual merger fractions, we calculate that the total number of mergers
a galaxy undergoes can be expressed as $N_{\rm merg} = 2.5 \tau_{m}^{-1}$.

Based on the changes in the measured merger fraction from Conselice
et al. 2009 (Paper III), Conselice et al. (2009b, in prep), the
most likely merger time-scale at $z < 1.2$ is $\tau_{\rm m} = 0.6\pm0.3$
Gyr.  Using this, we calculate that the total number of mergers
which occur at $z < 6$ is $N_{\rm merg} = 4.2^{+4.1}_{-1.4}$.  However,
between $z = 4 - 6$, the average number of mergers occurring is roughly 0.5, 
and 
thus not every galaxy, on average, will go through a merger during this epoch.
However, most of the merging within these massive galaxies occurs at 
$z > 1$, independent
of the value of the merger time-scale, as discussed earlier in Conselice 
et al. (2009).    We note that this is similar to the pair fraction history
for the most massive galaxies with M$_{*} > 10^{11}$ \solm found by
Bluck et al. (2009) between $z = 3$ and $z = 0$.

\begin{figure}
%\vspace{5.5cm}
 \vbox to 120mm{
\includegraphics[angle=0, width=85mm]{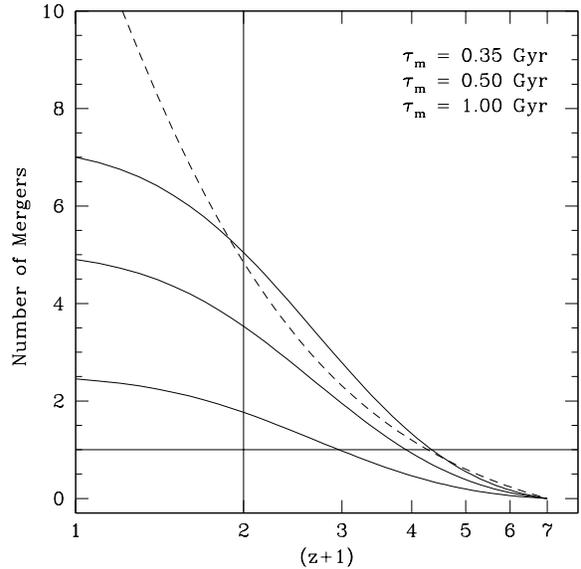}
 \caption{The integration of the inverse of $\Gamma$, which gives
the total number of mergers since $z = 6$, as a function of redshift.  
The three solid
lines show the evolution of the number of mergers that have occurred for
galaxies with M$_{*} > 10^{10}$ \solm since $z = 6$ using different
values for the time-scale in which the CAS system is sensitive
to merging (see text).  The dashed line shows the evolution for
mergers using a constant time-scale of $\Gamma = 1$ Gyr. As can be
seen from the comparison of the constant $\Gamma$ line, the evolution
of the merger rate declines rapidly at $z < 1$.  }
%\vspace{3cm}
} \label{sample-figure}
\end{figure} 

\subsection{Smooth and Possibly Relaxed Systems}

%old stuff
%fits for C vs. Radius
%bdrop: C(z) = 0.34+/-0.04*R_e - 0.12+/-0.11
%vdrop: C(z) = 0.53+/-0.16*R_e - 0.61+/-0.43
%idrop: C(z) = 0.65+/-0.21*R_e - 0.99+/-0.59

%holding zero point const. at -0.12
%vdrop: C(z) = 0.36+/-0.03
%idrop: C(z) = 0.35+/-0.02

\subsubsection{Relation of Size and Concentration}

As described briefly in \S 5.1, based on visual estimates of 
structure, we find that a significant fraction of our sample 
of drop-outs at $z > 4$ are smooth, and thus perhaps dynamically
relaxed 
systems. As described in \S 5.1 and shown in Figure~10, a large
fraction of our sample ($>50$\%) have smooth symmetrical
morphologies.  Examples of these galaxies, selected as $i$-drops, are 
shown in Figure~16.  Other evidence, besides being 
smooth and symmetrical, for these
systems as bound and perhaps `relaxed' after initial
formation is lacking.  Yet we have found a 
correlation between the concentration of light in the
observed $z-$band, and the sizes of these 
galaxies that possibly reveals this is the case.

\begin{figure}
%\vspace{5.5cm}
 \vbox to 85mm{
\includegraphics[angle=0, width=85mm]{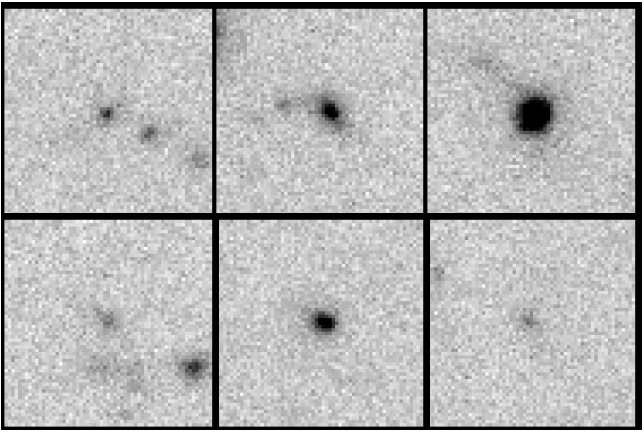}
 \caption{Images of galaxies at $z \sim 6$ ($i-$drops)
which have a smooth and symmetrical morphology and structure.
The system at the upper right is the galaxy `1ab' with
a spectroscopic redshift of $z = 5.8$ (Yan et al. 2005). The
field of view of each image is 1.8\arcsec\, on a side, or 
10 kpc at $z \sim 6$.}
%\vspace{3cm}
} \label{sample-figure}
\end{figure}

Before this study, the major known properties of $z > 4$ 
galaxies were the star formation 
rates, stellar masses, and the sizes of these systems (e.g., Ferguson et
al. 2004; Bouwens et al. 2006). What is generally found is that galaxies
are smaller in size, and have lower stellar masses at higher
redshifts.  What is not known is whether there are any scaling
relations between the various quantities for these high-redshift
galaxies. 

We have discovered what is perhaps the first fundamental
scaling relation for galaxies at $z > 4$, between the sizes
of our sample of LBGs, as measured through the half-light radius, and
the CAS concentration index.   What we find is that for systems
that are not asymmetric, or bimodal in structure, there is a relation
such that galaxies with a larger effective radius have a larger
concentration index.  We find that the quantitative scaling 
between the concentration index and the half-light radius is given by

\begin{equation}
R_{\rm h} = (0.33\pm0.04) \times C - (0.13\pm0.11),
\end{equation}

\begin{figure*}
%\vspace{5.5cm}
 \vbox to 90mm{
\includegraphics[angle=0, width=174mm]{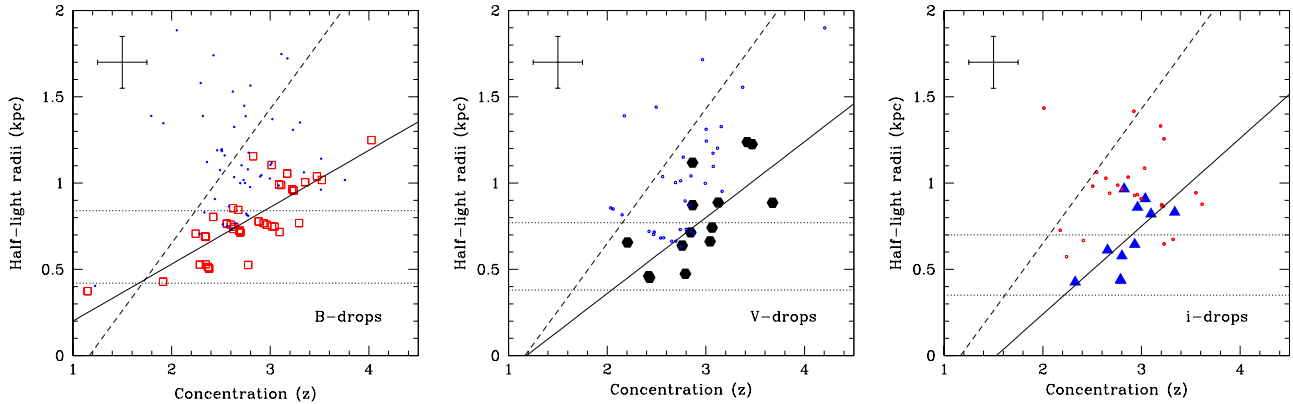}
 \caption{The relation between size (half-light radii) and 
light concentration for
our sample of UDF galaxies.  The open boxes, solid circles and
triangles are in the panels for the B-drops, V-drops and $i-$drops, 
respectively. 
Those galaxies in each of the drop-out bins that are either
asymmetric $A > 0.2$ or have a low light moment (M$_{20} < -1.5$) are
shown as dots.   The solid line is the best fit between $C$ and
half-light radii for each drop-out.  The dashed line shows the
best fit between $C$ and half-light radii for early-type galaxies
at $z < 1$ (Lanyon-Foster et al. 2009).  The two lines show the
size of the ACS camera PSF's FWHM and one half this value.
 There is clearly a strong relation between the
size and concentration index for these normal systems, which is
not found for those systems with a peculiar structure. }
\vspace{4cm}
} \label{sample-figure}
\end{figure*}
 
\noindent for the B drop-outs within our sample.  The
relationship between the $C$ parameter and the
half-light radius ($R_{\rm h}$) is shown in
Figure~17.  We also plot on Figure~17 our sample divided
up into two different classes. The first are the
smooth systems, selected by the criteria $A < 0.1$
and M$_{20} < -1$. These smooth galaxies are shown as
the open boxes on Figure~17, and are by definition 
those systems which are neither asymmetric nor
have multiple components, as shown through the
asymmetry and M$_{20}$ indices.  The small
dots on these graphs are for those systems
which do not meet the above asymmetry and M$_{20}$
criteria, or in other words have $A > 0.1$ and M$_{20} > -1$.  These
systems are asymmetric and show a distorted 
structure.

\begin{figure}
%\vspace{5.5cm}
 \vbox to 120mm{
\includegraphics[angle=0, width=85mm]{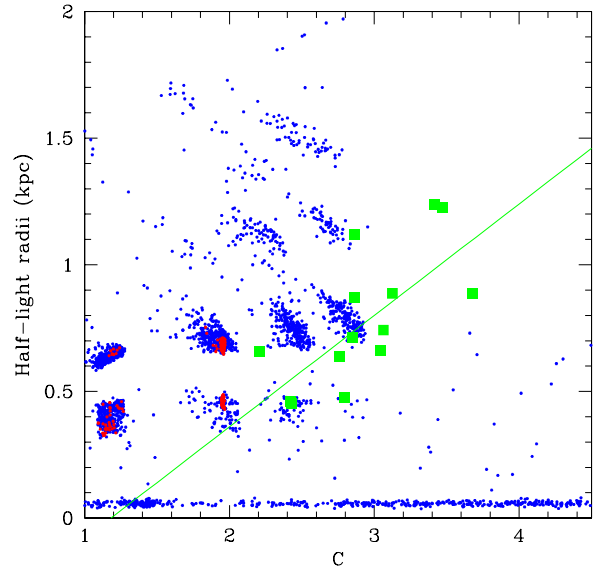}
 \caption{The relation between the half-light radius, assuming
all objects are at $z = 5$, vs. the concentration index.  The
blue points are faint galaxies within the COSMOS field, while
the red points include 3736 stars within the COSMOS field, nearly
all of which are at four different well defined locations in this
parameter space.  The green squares
are the relation between these two quantities for the V-drops,
while the solid line is the fit of these points (see Figure~17).  }
%\vspace{3cm}
} \label{sample-figure}
\end{figure} 

%\begin{figure*}
%\vspace{5.5cm}
% \vbox to 80mm{
%\includegraphics[angle=0, width=174mm]{conselice.figdono.eps}
% \caption{The relationship between the half-light radius,
%as measured in kpc using our cosmology, vs. the M$_{20}$
%index.  As in Figure~11, the red triangles are for those
%galaxies that are selected as mergers or by being peculiar in
%structure.  The horizontal dashed lines show the size of
%the ACS PSF before drizzling, and half that value.  
%The solid line shows the sized of the PSF in our drizzled
%images (see text). }
%\vspace{3cm}
%} \label{sample-figure}
%\end{figure*}

We likewise find similar scaling relations for
the V-drops and the i-drops as seen in the observed
$z-$band ACS imaging.  Quantitatively, the relation for
the V-drops is given by:

\begin{equation}
R_{\rm h} = (0.44\pm0.13) \times C - (0.52\pm0.36),
\end{equation}

\noindent while the relation for the i-drops is given by:

\begin{equation}
R_{\rm h} = (0.51\pm0.16) \times C - (0.78\pm0.46).
\end{equation}

%idrop: C(z) = 0.65+/-0.21*R_e - 0.99+/-0.59

\noindent These correlations are significant at
the $> 3 \sigma$ level.  We also plot on Figure~17, as a
dashed line, the relation between
the half-light radii and concentration index for elliptical
galaxies at $z < 1$ taken from GOODS imaging (Lanyon-Foster
et al. 2009 in prep).  Lanyon-Foster et al. (2009) present
a general study of how concentration and size correlate
for galaxies of different types at $z < 1$,
and shows that only normal early-type galaxies follow
a relation between $C$ and the half-light radii.

\begin{figure*}
%\vspace{5.5cm}
 \vbox to 90mm{
\includegraphics[angle=0, width=174mm]{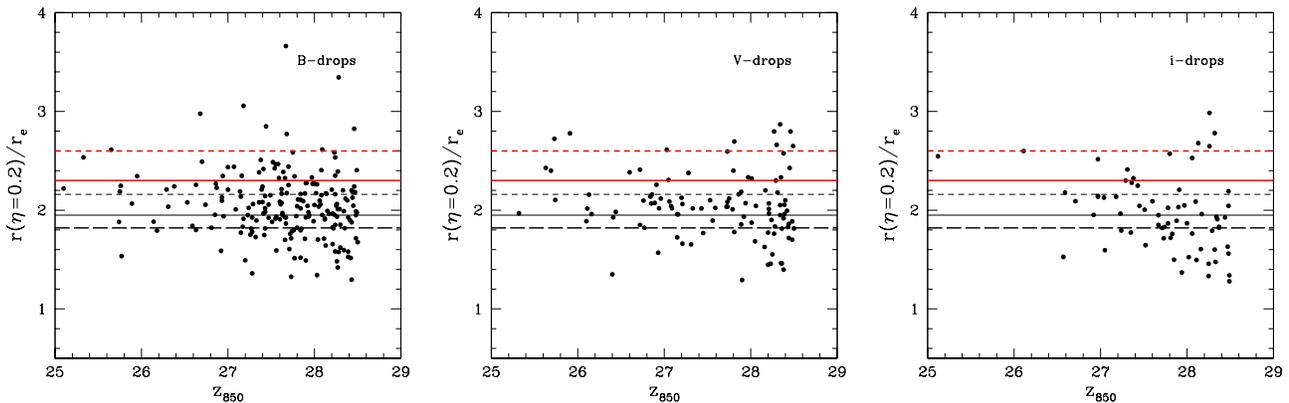}
 \caption{The ratio of the two radii we use in this paper, the
Petrosian radius, r($\eta = 0.2$), and the half-light radius, r$_{\rm e}$,
plotted as a function of magnitude.  The various horizontal lines show the
ratios for
these radii for both theoretical profiles, and empirical data.  The
black solid line show the ratio for a model Gaussian profile, the
short-dashed black line shows the ratio for an exponential profile,
while the long-dashed line shows the ratio for an ideal r$^{1/4}$ 
profile.  The thicker red lines show: empirical data from nearby
galaxies with blue colours (B-V) $< 0.85$ (solid), while the red dashed
line shows the ratio for redder galaxies with (B-V) $>0.85$ (see
Bershady et al. 2000). }
\vspace{4cm}
} \label{sample-figure}
\end{figure*}

\subsubsection{Reliability of Result}

This relation between the concentration and half-light radius,
which is generally seen for early-type galaxies at $z < 1$,
is such that some of our galaxies have measured half-light
radii sizes which are similar to the FWHM of the ACS
PSF as shown in Figure~17 by the upper horizontal line. As
described in \S 3.2.1, 3.3 and 4.1, this resolution is sufficient to
measure the sizes of these systems, as well as their
concentration values for most galaxies.  This is further
borne out by the consistency of the measured concentration
index as measured after simulating drop-out galaxies
to higher redshifts (\S 4.1).

Another way to address this issue is to determine how the
stars seen within the COSMOS field (\S 3.3) fall within the
concentration vs. half-light relation. As we have already
seen, the concentration values for these stars are all
at $C < 2$.   As a further test, in Figure~18 we plot the concentration vs.
half-light radii for our faint galaxy and star sample from
the COSMOS field, with the assumption that they are all
at $z = 5$, and thus mimicking our V-drop sample.  

Figure~18 demonstrates that the stars within the COSMOS
sample all fall into four different areas of the concentration
vs. size plane.  We note that none of the galaxies we study
have concentration values as low as these stars, and this is
further evidence that the PSF is not dominating the measured
concentrations.  Furthermore, we find that all of the half-light
diameter measurements are either just at, or larger than, the size
of the drizzled ACS PSF (comparable with the lower dashed line).   
It however remains possible that some of our
systems have sizes smaller than their measured values,
and our measurements are upper limits for galaxies
at $C < 3$.  Thus, while it
may not be the case that these galaxies follow a strong
linear relation between size and concentration, it is
unlikely that galaxies with smaller concentrations have
larger half-light radii than galaxies with higher concentrations.

We  can furthermore demonstrate, using the ratio of different radii, 
that there is a relationship
between light concentration and size, and that this is not
due to effects from the PSF.  We can get some idea about the light profile
shape for these systems through examining the ratio of the
Petrosian radius and the half-light radius.  We show the correlation
between this ratio and $z-$band magnitude in Figure~19.  Various
profile shapes have different ratios of r($\eta = 0.2$)/r$_{\rm e}$ $-$
with various empirical and theoretical ratios for this ratio shown.

As can be seen, our sample of galaxies span the range of possible
radii ratios, as well as have ratios similar to nearby elliptical and
spiral galaxies.  Also, we furthermore do not see that at fainter
magnitudes the ratio approaching one particular value, such as a
Gaussian.  There is a slight tendency for this ratio of radii
to approach smaller values at fainter magnitudes, but this
occurs below our $z = 27.5$ magnitude cut.

\subsubsection{Interpretation}

The meaning of the correlation between concentration and
size is likely related to the
fact that for nearby elliptical galaxies, the concentration index
correlates with the stellar mass (e.g., Conselice 2003).
This relation is such that galaxies with a larger
concentration index have a larger mass.  Several
examples of these smooth galaxies are shown in 
Figure~16. One of the brighter systems is shown in
the upper right, a galaxy called `1ab' by Yan et al.
(2005).  This system, at $z_{\rm spec} = 5.83$, 
is calculated by Yan et al. to contain a stellar mass of 
4.3 $\times 10^{10}$ \solm and has a stellar population 
age of 0.5 Gyr.  It is
therefore a fairly old massive galaxy.  It contains
a concentration index of $C = 3.5\pm0.2$, which places it at
the upper end of our C vs. half-light relation.

We argue that the correlation between the sizes of these
Lyman-break galaxies and their light concentrations suggests that
these galaxies are at least temporarily relaxed systems
that have either formed rapidly in a single burst, or
have had a merger some time ago. They are also the only systems
besides ellipticals at $z < 1$ which show a correlation
between size and concentration, suggesting that
the formation modes for these systems may be similar
to those of ellipticals.  However, the sizes of these
systems, at a given concentration, 
are smaller by a factor of $> 2$ compared with $z < 1$
systems (e.g., Buitrago et al. 2008).  
This difference is either due to
an intrinsic growth after multiple mergers (e.g., Trujillo
et al. 2007) or we are missing the outer parts of these
systems.  Massive ellipticals at $z > 1$ are found to be more compact
and smaller (e.g., Trujillo et al. 2007; Buitrago et al. 2008),
than those at $z = 0$, and it is possible
that these systems are the initial formation of early-type galaxies.

It appears therefore that these systems are in, at least,
a temporarily relaxed state.  By using the results of 
Section 5.1.2, we can conclude that these galaxies must have
had their last major dynamical assembly episode at least
0.5 Gyr earlier, and perhaps even 1 Gyr.   
A time-scale of 0.5-1 Gyr is similar to the age of the universe
for these redshifts, particularly and obviously for the $i-$drops.  
It is possible that the merger signatures of assembly have
dissipated by 0.5 Gyr, although this would imply that the last
merger occurred at $z > 10$.  If we are able to see merger
signature for 1 Gyr, then we could rule out any merging activity
occurring for at least the smooth $i-$drop sample.   This implies
that at least some of these systems  were
not formed by a major merger process, but have had an assembly produced
through a rapid collapse of gas, or a rapid assembly through the
accretion of gas (e.g., Keres et al. 2005).  The merger process 
however can be responsible
for some of the B$-$drops and V$-$drops, although relaxation
time-scales are likely too long to account for the $i-$drops.

This time-scale of $\sim 1$ Gyr is furthermore similar to the time-scale
between mergers found by the $\Gamma$ index, described in
\S 5.3 at $z > 1$.  This indicates that the two methods for measuring
the time-scales for merging, which are independent, reveal that
the time-scale for merging and relaxation are similar. We discuss
the implications of this, and what it reveals about the
structure formation of early galaxies, in \S 6.

\subsection{Relation to Star Formation}

We use the observed ultraviolet flux (1250-2500 \AA) from our 
data to determine the unobscured star formation rate within
our sample of Lyman-break galaxies.  To measure the star formation
rate we use the relation between UV flux and the ongoing
star formation rate, as derived by Kennicutt (1998) and references
therein,

\begin{equation}
{\rm SFR\, (M_{\odot}\, yr^{-1})} = 1.4 \times 10^{-28}\,{\rm L_{\nu}} ({\rm erg\, s^{-1} Hz^{-1}})
\end{equation}

\noindent which assumes a Salpeter IMF.  This is very similar to previous
relations used by e.g., Madau, Pozzetti \& Dickinson (1998).
The reliability of this equation is
uncertain, and at best it is accurate to within a factor of 2-3.  For
example, using
an IMF such as Scalo would produce star formation rates a factor of
$\sim 2$ higher.  It is 
even more uncertain when trying to understand the total star formation rate 
within a galaxy due to unknown dust corrections.  This later effect may not 
be a 
major issue, as by definition our LBGs are UV bright, and thus cannot be dust
dominated as shown by Adelberger \& Steidel (2000).

With these caveats, we determine the relationship between the star formation
rate of our galaxies and their structural features.  Previous similar
studies at lower redshifts have found little or no correlation between
structural features, as measured in the ultraviolet, and the star formation 
rate or other physical features (e.g., Law et al. 2007; Peter et al. 2007).
However, by using optical morphologies there are correlations between the
structures of galaxies and the underlying physical properties (e.g.,
Conselice 2003; Conselice et al. 2005).  The rest-frame UV CAS parameters have
also never been examined in terms of the measured star formation
rate.  In general, the only parameter that strongly correlates with
star formation in the nearby universe is the clumpiness index, $S$.

\begin{figure*}
%\vspace{5.5cm}
 \vbox to 160mm{ 
\includegraphics[angle=0, width=174mm]{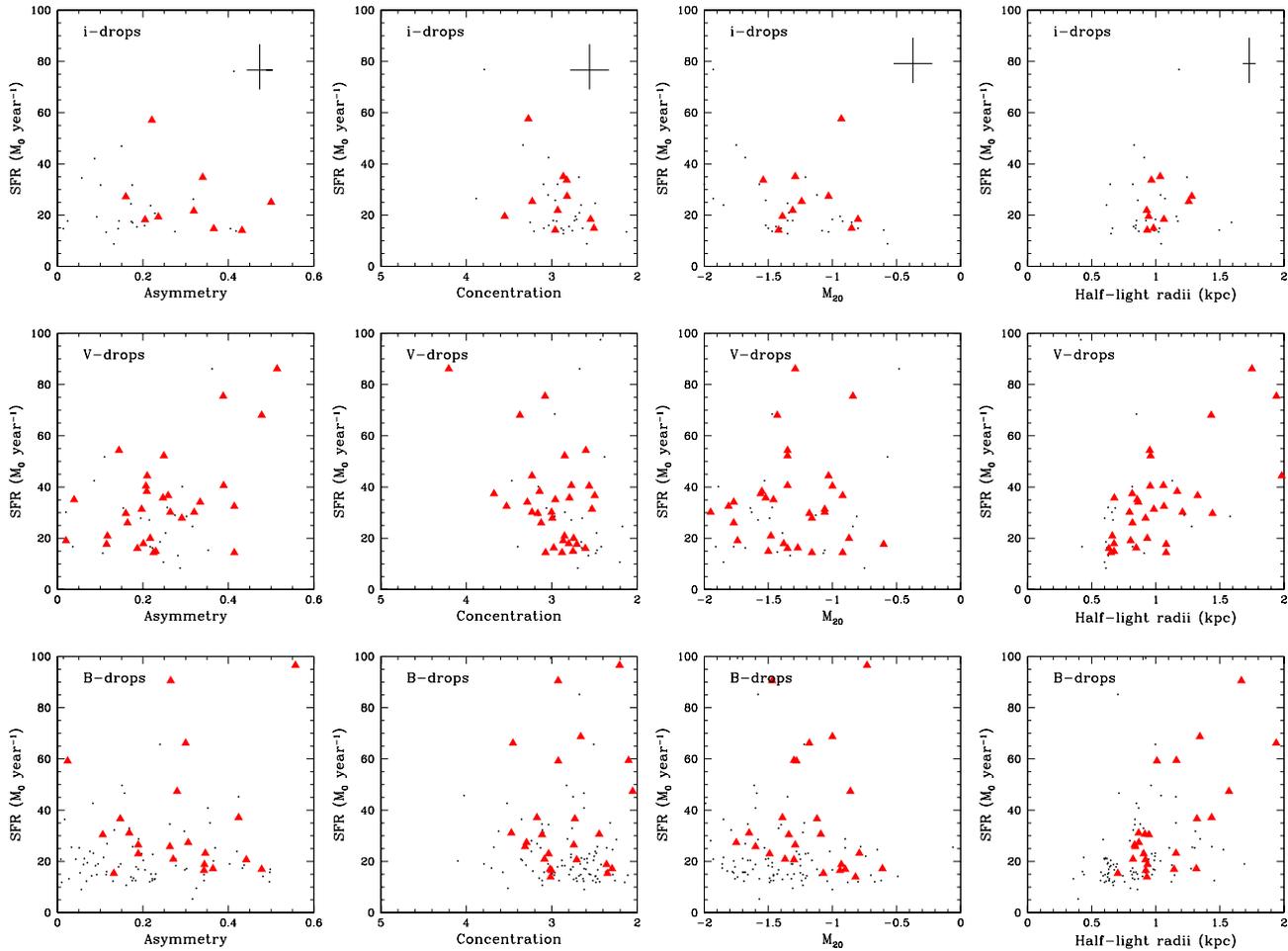}
 \caption{The relation between the ongoing unobscured star formation
rate, as measured using rest-frame UV light (\S 5.5) and
the asymmetry, concentration, M$_{20}$ index, and the
half-light radii (kpc).  As in earlier figures, the
triangles show those systems which appear by eye to
be distorted, or in a merger phase, while
the dots are those systems which appear smoothed and
possibly dynamically relaxed.   The typical error bars
for our measurements are at the top of the upper panels.}
%\vspace{3cm}
} \label{sample-figure}
\end{figure*}

With a few important exceptions, we find that within our sample there is 
very little
to no correlation between structural parameters and star formation
rates as measured by equation (17).  We plot these correlations on Figure~20,
which shows how star formation relates to the concentration index,
the asymmetry index, the M$_{20}$ values, and the 
half-light radii (r$_{e}$), with
the unextincted star formation rate.  There are a few slight correlations 
which can be seen.  

One correlation is that for some drop-outs, there
is a slightly higher star formation rate for those  which appear 
distorted or peculiar.  We find that within the $i$-drops the
star formation rate for the galaxies classified as peculiars is
$\Psi$ = 27$\pm13$ \solm year$^{-1}$, while the $i-$drops classified as normal
have a star formation rate of $\Psi$ = 28$\pm30$ \solm year$^{-1}$. 
The star formation rate for the $i-$drops thus does not appear to depend
on the apparent visual morphology of the system.  At lower redshifts, there
is a larger difference, with the distorted galaxies revealing
a higher star formation rate.  The V-drops which are peculiar have
a star formation rate of $\Psi$ = 63$\pm139$ \solm year$^{-1}$, with
the average dominated by a few very highly star forming systems.  The
normal V-drops have a star formation rate of 
$\Psi$ = 28.0$\pm30$ \solm year$^{-1}$.  Likewise for the lowest
redshifts systems in our sample, the B-drops, the star formation
rate for the peculiar systems is $\Psi$ = 49$\pm43$ \solm year$^{-1}$,
while for the normal galaxies it is $\Psi =$ 24$\pm17$ \solm year$^{-1}$.

%0 = nothing
%1 = v. faint
%2 = noisy
%3 = star-like
%4 = elongated
%5 = asymmetric
%6 = normal
%7 = neighbor
%8 = unusual/peculiar
%9 = star=trail
%10 = off=frame

Another correlation is that, on average, systems
which are more concentrated, as measured by both the concentration and
M$_{20}$ indices, have a higher degree of star formation.  This probability 
is such that by using a generalised Kendall's Tau there is
a $\sim$ 0.04 probability  that a correlation is not present between
star formation and concentration and M$_{20}$. There is 
also a similar correlation between the size of these galaxies
and their star formation rates.  This correlation is partially due
to larger galaxies having more area for star formation as this correlation,
as well as the others with size, is largely removed after comparing the 
star formation rate density, that is the star formation rate per unit
area, with the same parameters. We do not see any correlation
between star formation rate and the asymmetry parameter, which suggests
that in the different phases of formation, the star formation rate
remains similar.  However, the more concentrated, and likely more massive,
systems contain a higher star formation rate then less concentrated
galaxies reflecting the likely rapid assembly of these systems.

\section{Discussion}

\subsection{Overview of Results}

The major result from this paper is that the structures of the earliest
galaxies, as probed in ultraviolet light, are diverse. This
can be seen by examining the images of these systems within the
Hubble Ultra Deep Field (e.g., Figure~3-4 \& 16) at $z = 4 - 6$. Not
surprisingly, we do not find the same kinds of galaxies at 
lower redshifts, even at $1 < z < 2$ (e.g., Conselice et al. 2005).
In particular we see no spiral-type systems, or any obvious disk
galaxies, even in formation, although systems like this are seen
at $z < 2$ (Conselice et al. 2004).  We do however find that a
significant fraction of galaxies at $z > 4$ are peculiar and distorted, 
although
over half of these systems are smooth and symmetrical. We 
argue below that this diversity in appearance is 
the result of galaxy formation processes, and are not arbitrary.
We furthermore make the case that the galaxies which 
appear smooth and symmetrical are likely to be in a relaxed phase, while those 
which are asymmetrical are in an active phase of assembly, possibly due to
recent merging activity.      We furthermore make the case that we are
witnessing the earliest phases of galaxy evolution and we are able
to put constraints on what fraction of the first massive galaxies were
perhaps formed through a very rapid collapse of gas, as opposed to being formed
from mergers of lower mass galaxies at even earlier times.

\subsubsection{The Assembly of Galaxies at $z > 4$: Are Asymmetric LBGs Mergers?}

The question we address in this section is what type of formation modes the
asymmetrical, distorted galaxies in our sample are undergoing, and whether they
are fundamentally different from the smooth and symmetrical systems.  In
particular, we want to understand if the smooth and asymmetrical galaxies
are related to each other.  
The data and plots presented in this paper allow us to directly address this
question. First, there are two important observations which suggest that the
distorted and asymmetric galaxies are undergoing an assembly phase,
perhaps driven by mergers.

The first question we need to address is whether or not the third of
the LBGs we see in the UDF which are asymmetric are ongoing mergers of
some type.  They are certainly in an active phase of evolution due 
to their high star formation rates (\S 5.5), and due to
the fact that between 20-40\% of the galaxies we examine are peculiar
in structure, or have a large enough asymmetry to be considered a merger
within the CAS system (e.g., Conselice 2003; Conselice et al. 2008).  
Without kinematic measures for these systems, which are likely at least a 
decade away
if not longer, we cannot confirm with 100\% certainty that any one of
these systems are mergers.  Although, even with resolved spectroscopy, 
determining if galaxies are mergers is ambiguous (e.g.,
Law et al. 2007). 

However, what is clear is that these galaxies are in an assembly
phase, and are hosting ongoing star formation.  What is also clear,
especially from NICMOS observations of slightly lower redshift
Lyman-break galaxies (e.g., Conselice et al. 2005) is that the asymmetric
features in these galaxies are not produced by small-scale
features, such as star forming knots, but are bulk structures.
These bulk asymmetric structures do not have to be the result
of the merging
process, but they are due to an assembly process, perhaps cold gas 
accretion or minor mergers.  This is due to the fact that the
sizes of asymmetric galaxies are significantly larger than smooth
symmetrical galaxies (\S 5.4).  This is expected if the
asymmetric galaxies are forming by the merging of two or more
galaxies.

Despite this, we can make the case that a significant fraction of
the Lyman-break galaxies are indeed within a major merger phase.  The
reason for this is that roughly 20\% of the drop-out galaxies
within our sample are found within a galaxy pair.  This is just
smaller than the galaxy merger fraction for our systems from CAS
which are roughly f$_{\rm gm} \sim 0.35$ (\S 5.2).  This
ratio of pair to structural fractions is roughly what is
expected based on N-body simulations (e.g., Lotz et al. 2008b),
and is similar to what is seen for $z < 1$ merging 
galaxies (e.g., Paper III).  In summary, it is therefore inescapable
that 20\% of our systems at $4 < z < 6$ are involved in some
type of merger.   Overall, we find that on average, a massive galaxy with
M$_{*} > 10^{10}$ \solm will undergo 2.5 to 7 mergers at $0.2 < z < 6$,
depending on the  time-scale for finding mergers
within the CAS method (e.g., Conselice 2006; Lotz et al. 2008b), but
hardly any of this merging occurs at $z > 4$.

This merger history does not vary significantly within $z = 4-6$, 
and remains flat for both the pair and the 
structurally derived mergers (Figure~10).  Furthermore,
we show in Figure~15, and discuss in \S 5.3.3, that these
LBGs only undergo on average half a merger between
$z = 6$ and $z = 4$.  However, this is based on a
merger time-scale of $\tau_{\rm m} = 0.5$ Gyr, although it 
is possible
that multiple mergers are ongoing through this time-period.
If indeed these galaxies are merging within our best
estimate of the merger time-scale, then merging is not
the dominant method for adding mass to galaxies 
between $z = 4$ and 6, and in fact the mergers we
see within the $z = 4$ LBG population could have
started at $z = 6$ or earlier.   Furthermore, smooth
gas accretion could significantly add gas and stellar
mass to these galaxies (e.g., Keres et al. 2005). Perhaps the most
remarkable result from this paper is the fact that
more than half of our galaxy sample appear to be
smooth relaxed systems, even within their rest-frame
ultraviolet structure (\S 5.4).

\subsubsection{Symmetrical Galaxies - Very Rapid Collapses?}

We have presented evidence that a significant fraction of our
sample of LBGs within the Hubble UDF are smooth and possibly
relaxed systems.  Perhaps the major evidence is the fact that these
smooth systems, as defined by the asymmetry and M$_{20}$ indices,
show a significant correlation between their measured
half-light radii and the concentration of light. This correlation
is such that galaxies with higher concentration indices have
larger sizes.  There is no correlation (at $> 5$ $\sigma$ confidence)
between the concentration and size for galaxies which are asymmetric,
or show multiple components in their structure (\S 5.4).  This
is strong evidence that these two types of galaxies are from
different populations. We
have further argued that this relation is not an artifact of
our smaller galaxies having half-light radii similar to the size
of the ACS PSF.

We conclude that the systems which display a tight correlation
between size and concentration are those that are, at least temporarily,
in a relaxed state, while those which are distorted and asymmetric
are currently undergoing a dynamically assembly phase or merger.  This
implies that the smooth galaxies have not undergone a significant
dynamical event some time in the recent past.  
We further calculate, based on the likely internal velocities for these
galaxies, what the time-scale is for these smooth and symmetrical systems
to have been relaxed.  We find
that this time-scale, based on LBG sizes and likely masses, is roughly 
0.5 Gyr.
This is similar to the separation in time between mergers at $z < 2$, which
we calculate with the $\Gamma$ index, described in \S 5.3.3.  These
two methods give similar results - that is the time-scale between
two successive mergers, and the time for relaxation are similar, suggesting
that there is enough time between successive merging events for
these systems to become relaxed enough to appear smooth and symmetric 
at lower redshifts.

The time scales for relaxation are older than, or similar to, the age of 
the universe at $z > 4$.  This implies that most if not all of the smooth
galaxies we examine were likely formed very early, or within 
a mode where star formation can occur in-situ within a small area
or region over a short time period, such as cold gas accretion 
(e.g., Keres et al. 2005).  We however cannot rule out that
an assembly event occurred for these galaxies at $z > 10$, due to the
difficulty of finding merger signatures after 0.5 Gyr.  Thus,
to solidly determine initial galaxy formation will require
observations of $z > 10$ galaxies.  Because these galaxies are resolved,
it is unlikely that any bulk large-scale features will be seen when
these galaxies are imaged at higher resolution.  

This implies that these galaxies might be primordial in the sense
that they were not formed by the mergers of two pre-existing
galaxies. It is possible that these systems, which dominate our
LBG population at $z > 3$ are forming through the gradual smooth
accretion of intergalactic gas, as proposed by e.g., Keres et al.
(2005). If they were produced through a merger, we would
still see residuals from this process in a distorted structure.
Therefore we conclude that many LBGs galaxies, with stellar
masses up to M$_{*} = 10^{11}$ \solm, are not being formed by
mergers at $z > 4$, but are formed in a type of initial very rapid 
smooth formation at $z > 10$.  However, as we argued in this paper, 
and in previous work cited
throughout, the merger process is important for
building up the mass of massive galaxies down to $z = 1$, and is
most important during $1 < z < 3$ (e.g., Paper I; Bluck et al. 2009).

\subsubsection{Comparison to Models and Stellar Populations}

The question we would like to address is how to fit these results into
a framework or model for how galaxies form and evolve.  While
we now have a picture observationally for how galaxies form, we
would ideally like to understand these observations in terms of
the physics of galaxy formation, especially within a cosmological
context.  There are
currently very few models for how the stars in galaxies are distributed
at high redshift.  Predicting structures, asymmetries, concentrations,
and even the sizes of these galaxies is very difficult, and has not
yet been done to any satisfying degree, and certainly not at the
level in which we can compare our results with.  We can,
however, conclude that what we are likely witnessing is a formation
mode which repeats itself every few Gyr within galaxies. The
fact that there are so many smooth, and apparently dynamically
relaxed systems, suggests this is the case.  This is also borne
out by the time-scales for merging based on the $\Gamma$ factor
which traces the time between when a galaxy will undergo a merger.  

What we are perhaps witnessing at $z > 1.5$ is the early formation modes of
the Hubble sequence where galaxies alternate between dynamical assembly 
and/or mergers
and dynamically quiescent systems.  It is not clear if all of the smooth
systems will eventually merge again, or passively evolve, but it
seems likely that many of them will, given that at $z > 4$ an average
LBG in our sample will undergo a major merger every 2 Gyr.  This
structure formation for galaxies appear to continue down to lower
redshifts until $z < 1$ when disk galaxies become more
common (e.g., Conselice et al. 2005; 2008).  These observations
are consistent with the idea that  merging is an important 
process for the formation of galaxies in the early universe up until
when the universe was roughly half its current age.

\section{Summary}

We present in this paper the first systematic study of the structures
of galaxies at $z > 4$.   We  find a 
diversity in galaxy structure as seen in the observed ACS
$z-$band imaging of these systems. We find that there are
significant correlations between the structures of these
galaxies and other physical properties that suggest how these
systems are forming.  One caveat about our results is that
we are studying these galaxies in the rest-frame UV using
observed optical light from ACS on the Hubble Space Telescope.

  We find  that
roughly half of all the LBGs in our sample are distorted, or measured
to have large asymmetries, but that the remainder are smooth, and
apparently dynamically quiescent systems.
We infer that a large fraction of the distorted systems are undergoing a 
merger, or some type of assembly, based on their structures and the fraction 
of systems in pairs with
another drop-out.  We find that the pair fraction 
is very similar to the inferred merger fraction as measured through 
the CAS system.

We conclude that the distorted systems are those assembling, possible
through mergers,
while the smooth and symmetric systems are in a temporary
relaxed phase.  We in fact discover a remarkable correlation between
the light concentration of non-asymmetric, non-peculiar objects,
and the half-light radii of these systems.   This
correlation does not exist for asymmetric galaxies, which generally
shows a large scatter in sizes at each concentration. 
 This  suggests
that the symmetrical galaxies are not currently undergoing a merger, and
were thus not formed through the
merger process. Based on the sizes, and likely internal velocities for
these systems, we calculate that these galaxies formed $\sim$ 0.5 Gyr
before we observe them. This implies that some of these galaxies'
initial formation must be nearly as old as the universe itself, although
we cannot rule out merger events at $z > 10$.

We calculate time-scales for the merging
process, and find that massive galaxies with M$_{*} > 10^{9-10}$ \solm
undergo a merger every 1-2 Gyr at $z > 2$.  By integrating the
galaxy merger rate per galaxy, between $z = 6$ and $z = 0$ we infer
that between 2.5-7 mergers occur for massive galaxies at $z < 6$. Most 
of this merging occurs at $z > 1$, and by $z = 2-3$, every massive galaxy 
has undergone at least a single merger.  To make further progress in our
understanding
the role of mergers in galaxy formation will require a better
knowledge of the time-scales for merging (see Conselice 2009;
Paper IV).  This, rather than
measures of the merger fraction, is the limiting aspect for
deciphering how mergers are driving the evolution of massive
galaxies.  Our best estimate, using the merger time-scale
derived in Paper IV, is that the total number of mergers
for M$_{*} > 10^{9-10}$ \solm galaxies at $z < 6$ is 4.2$^{+4.1}_{-1.4}$.

We finally investigate how the merger properties of our drop-outs
relate to the ongoing star formation rate, as measured through the
UV light emitted from the same systems. We generally find no strong
correlation between the star formation rate and the CAS or
size parameters, although there is an indication that more
concentrated and larger galaxies have higher star formation
rates.  We also find a tentative higher star formation rate for
distorted galaxies, in comparison to smoother systems.

Finally, we show that our smooth galaxies have a formation time-scale similar
to the merger time-scale, and that what we might be seeing is 
a population of galaxies in ongoing or post-merging activity.  Our
merger rate calculations suggest that up to 2 Gyr occurs between
merging events for these systems, allowing them time to become
smooth relaxed systems, which we can see in the correlation between
size and concentration index.   This merging continues down to
$z \sim 1-1.5$, as seen in several other papers, including paper
I of this series (Conselice et al. 2008).  After this,
 disk galaxies become common and
major merging ends as a dominate assembly method for forming the masses of
galaxies.  

While these observations probe nearly the beginning of galaxy assembly,
there is still the possibility that some even early initial formation
events 0.5 Gyr earlier than $z = 6$ occurred.  Making further progress
on the initial formation of galaxies will require resolve imaging of
$z > 10$ galaxies.  Because the WFC3 on Hubble will have a courser resolution
than ACS, it will be difficult to examine higher redshift galaxies in similar
ways, even if a suitable $z > 10$ population is identified.
It is likely that JWST or ground-based adaptive optics of selected sources
will be required to probe the earlier phases of galaxy formation, utilising
structures, than what we have examined in this paper.

  Support
for this work was provided by a Summer Undergraduate Research Fellowship
(SURF) from the California Institute of Technology, and support 
from the UK Science and Technology Facilities Council (STFC).  We thank
the Hubble Ultra Deep Field team for making their data products readily
available. We thank Asa Bluck for several insightful comments on this work 
and the resulting
paper, and Norman Grogin and Chien Peng for discussions of the ACS PSF.

\appendix

\label{lastpage}

\end{document}